\documentclass[acmsmall]{acmart}  

\AtBeginDocument{%
  }

\usepackage{dsfont}
\usepackage{threeparttable}
\usepackage{amsmath}
\usepackage{amsfonts}
\usepackage{color}
\usepackage{comment}
\usepackage{graphicx}
\usepackage{multirow}
\usepackage{bm}
\usepackage{url}
\usepackage{booktabs}  
\usepackage{enumitem}
\usepackage{float}
\usepackage{array}
\usepackage{float}
\usepackage{subfig}

\usepackage{algorithm}
\usepackage{algorithmic}

\begin{document}

\title{Cluster-based Graph Collaborative Filtering}

\author{Fan Liu}
\affiliation{%
  \institution{National University of Singapore}
  \streetaddress{21 Lower Kent Ridge Road
}
  \city{Singapore}
  \country{Singapore}
  \postcode{119077}
}
\email{liufancs@gmail.com}

\author{Shuai Zhao}
\email{zhao254014@163.com}
\affiliation{%
  \institution{Qilu University of Technology (Shandong Academy of Sciences)}
  \streetaddress{Shandong Artificial Intelligence Institute, 19 Keyuan Road}
  \city{Jinan}
  \state{Shandong}
  \country{China}
  \postcode{250014}
  }

\author{Zhiyong Cheng}
\email{jason.zy.cheng@gmail.com}
\affiliation{%
  \institution{Hefei University of Technology}
  \streetaddress{193 Tun Xi Lu, Baohe District}
  \city{Hefei}
  \postcode{230002}
  \country{China}
  }

\author{Liqiang Nie}
\affiliation{%
  \institution{Harbin Institute of Technology, Shenzhen}
  \streetaddress{School of Computer Science and Technology, HIT Campus of University Town of Shenzhen}
  \city{Shenzhen}
  \country{China}
  \postcode{518055}
}
\email{nieliqiang@gmail.com}

\author{Mohan Kankanhalli}
\affiliation{%
 \institution{National University of Singapore}
 \streetaddress{School of Computing, 13 Computing Drive}
 \city{Singapore}
 \country{Singapore}
 \postcode{117417}
 }
 \email{mohan@comp.nus.edu.sg}


\begin{abstract}
Graph Convolution Networks (GCNs) have significantly succeeded in learning user and item representations for recommendation systems. The core of their efficacy is the ability to explicitly exploit the collaborative signals from both the first- and high-order neighboring nodes. However, most existing GCN-based methods overlook the multiple interests of users while performing high-order graph convolution. Thus, the noisy information from unreliable neighbor nodes ($e.g.$, users with dissimilar interests) negatively impacts the representation learning of the target node. Additionally, conducting graph convolution operations without differentiating high-order neighbors suffers the over-smoothing issue when stacking more layers, resulting in performance degradation.
In this paper, we aim to capture more valuable information from high-order neighboring nodes while avoiding noise for better representation learning of the target node. To achieve this goal, we propose a novel GCN-based recommendation model, termed Cluster-based Graph Collaborative Filtering (ClusterGCF). This model performs high-order graph convolution on cluster-specific graphs, which are constructed by capturing the multiple interests of users and identifying the common interests among them. Specifically, we design an unsupervised and optimizable soft node clustering approach to classify user and item nodes into multiple clusters. Based on the soft node clustering results and the topology of the user-item interaction graph, we assign the nodes with probabilities for different clusters to construct the cluster-specific graphs. To evaluate the effectiveness of ClusterGCF, we conducted extensive experiments on four publicly available datasets. Experimental results demonstrate that our model can significantly improve recommendation performance.
\end{abstract}

\ccsdesc[500]{Information systems~Personalization}
\ccsdesc[500]{Information systems~Recommender systems}
\ccsdesc[500]{Information systems~Collaborative filtering}

\keywords{Collaborative filtering, Recommendation, Graph Convolutional Network, Clustering, Multiple Interests}


\maketitle

\section{Introduction}
Recommendation systems have become essential for various online platforms. It not only assists users in navigating through overwhelming volumes of information but also serves as a powerful tool for service providers to increase customer retention and revenue. Within this field, Collaborative Filtering (CF)-based models have achieved significant advancements in learning the user and item representations by effectively exploiting the historical user-item interactions~\cite{netflix,he2017neural,wu2016cdae,xue2017deep}. For instance, Matrix Factorization (MF) decomposes the user-item interaction matrices into lower-dimensional latent vectors and predicts the user preference with the inner product~\cite{netflix}. By utilizing nonlinear neural networks, NeuMF~\cite{he2017neural} can model the complex interaction behaviors, yielding better recommendation performance. 
More recently, Graph Convolution Networks (GCNs)-based models~\cite{berg2019gcmc,wang2019ngcf,wang2019kdd,He@LightGCN,liu2020A2GCN,wang2020DGCF,Liu2021IMP_GCN,fan2022GTN,Wei2023LightGT} offer a cutting-edge approach in recommendation systems by leveraging high-order connectivities based on the topological structure of the interaction data. These models obtain the final node representations ($a.k.a$ user and item embeddings) by iteratively aggregating information from neighboring nodes. Many methods have demonstrated that exploring additional information based on graph topology structure can efficiently enhance user and item representation, ultimately improving recommendation performance. For example, NGCF~\cite{wang2019ngcf} incorporates high-order connectivity of
user-item interactions in the user-item bipartite graph for representation learning in recommendation. With careful experimental studies, He et al.~\cite{He@LightGCN} simplified NGCF and achieved better performance by removing the feature transformation and the non-linear activation module. 
This is because these two components do not contribute positively to the final performance, and may even have a negative effect due to their impact on training difficulty.

Despite the notable success of GCNs in recommendation systems, two key challenges persist in the information aggregation process of neighboring nodes. 
Firstly, \textbf{noisy information from unreliable neighboring nodes harms the representation of the target nodes}~\cite{Liu2021IMP_GCN,fan2022GTN}.
Most existing GCN-based recommendation models are based on a user-item graph where interactions link users and items. They generally assume that embedding learning could take advantage of collaborative signals from high-order neighbors. 
Thus, they aggregate the messages passing from neighboring nodes to form the target user/item representation without distinguishing the high-order neighbors. As a result, the noisy information from unreliable users and items is also involved in the representation learning of a target node, resulting in sub-optimal performance. For instance, user $u_1$ has a strong preference for mystery novels. However, a high-order neighboring user $u_2$, who is loosely connected to $u_1$, prefers gardening books. In GCN-based models, this noisy information about gardening books could be delivered to the representation of $u_1$. Some efforts have been made to enhance the quality of information delivered from the neighbor nodes. For example, IMP-GCN~\cite{Liu2021IMP_GCN} devises an interest-aware strategy to only pass messages from user nodes with common interest and their associated item nodes when performing high-order graph convolution. 
To capture the adaptive reliability of the interactions between users and items, GTN~\cite{fan2022GTN} introduces a graph trend collaborative filtering technique to achieve the desired adaptive smoothness property for the interaction graph.
Secondly, \textbf{GCN-based recommendation methods suffer from the over-smoothing problem}~\cite{wang2019ngcf,ChenWHZW20,He@LightGCN,Liu2021IMP_GCN}. Most GCN-based recommendation models achieve their peak performance by stacking only a few layers, typically 2 or 3. Increasing the number of layers leads to a significant drop in performance, as the graph convolution operation smooths the graph Laplacian and makes node representations indistinguishable after multiple convolutions~\cite{Liu2021IMP_GCN}.
For example, Chen et al.~\cite{ChenWHZW20} found that stacking more layers leads to over-smoothing, causing user and item representations to become more similar in recommendation. To alleviate this problem, they simplified the network structure by removing the non-linearities and introduced a residual network structure, which significantly improved the accuracy of recommendations.
To maintain the uniqueness of the users by stacking more graph convolution layers, Liu et al.~\cite{Liu2021IMP_GCN} grouped users based on their shared interests and performed high-order graph convolution on subgraphs. Furthermore, as demonstrated in \cite{CommuDetection2022RS}, detecting the users within each community who share similar tastes can improve recommendation accuracy.
\begin{figure}[t]
	\centering
	\subfloat[User Nodes]{\includegraphics[width=1.8in]{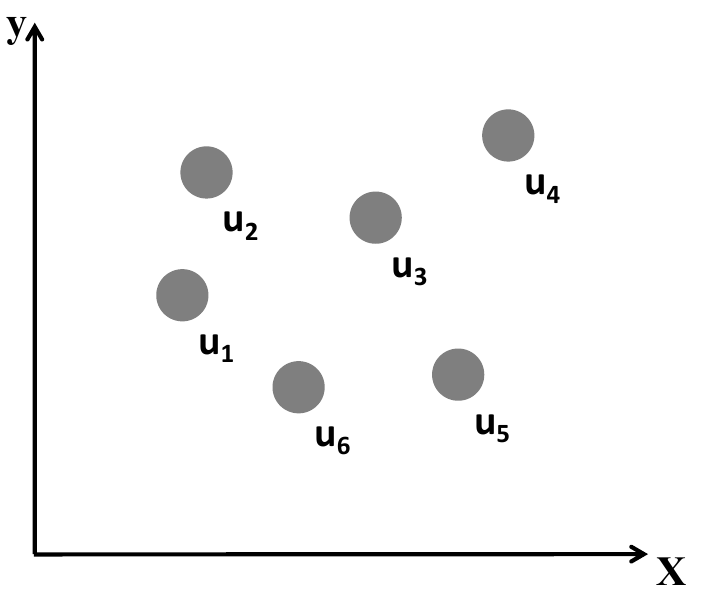}}
	\subfloat[Hard Clustering]{\includegraphics[width=1.8in]{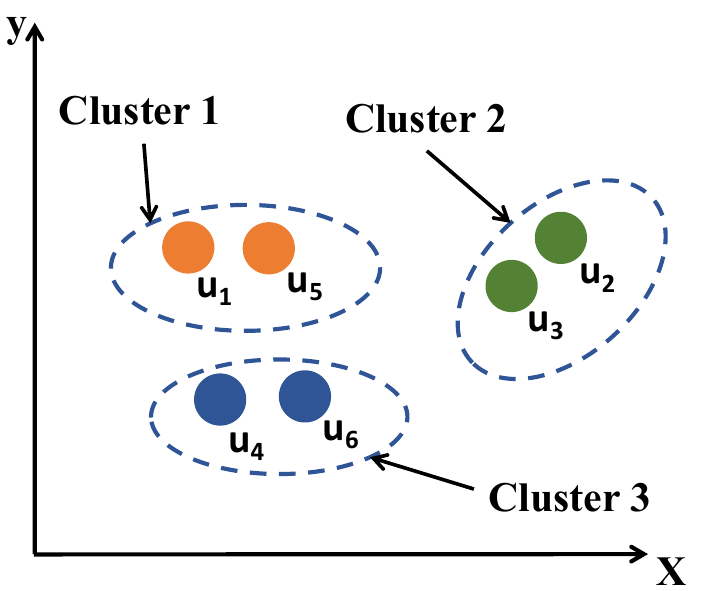}}
	\subfloat[Soft Clustering]{\includegraphics[width=1.8in]{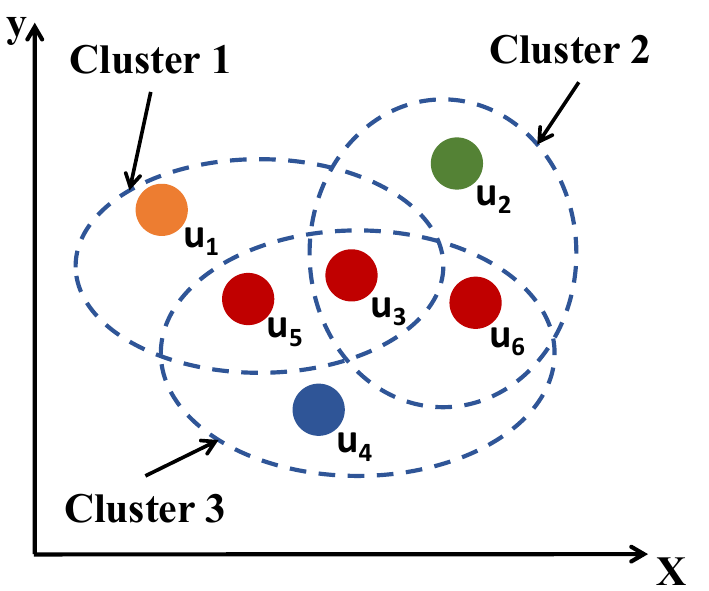}}
	\vspace{0pt}
	\caption{An example of user clustering. (a) A toy example containing 6 user nodes. (b) Hard clustering. Every user node can belong to only one cluster. (c) Soft clustering. Every user node may belong to several clusters with a fractional degree of membership.}
	\vspace{0pt}
	\label{fig:soft-cluster}
\end{figure}


Given these challenges, previous works generally ignore high-order connections within the graph, overlooking users' multiple interests in high-order graph convolution operations.
Indeed, in ~\cite{Liu2021IMP_GCN}, the authors demonstrated that users receive more valuable information from other users with common interests and their associated items. As illustrated in Fig.~\ref{fig:soft-cluster}(a) and (b), to identify the users with common interests, they classify users into different groups using a hard clustering method. However, users often exhibit diverse and overlapping interests across a range of topics or item categories (e.g., movies of different genres).
For example, a user might prefer both science fiction and romantic movies. Similarly, a movie is probably categorized into both science fiction and action. The hard clustering fails to capture this overlap, resulting in limited utilization of information. In contrast, with soft clustering, a user could be classified into multiple clusters representing different interests.
As shown in Fig.~\ref{fig:soft-cluster}(c), user $u_3$ has a common interest with user $u_1$ in cluster 1 while also sharing an interest with user $u_2$ within cluster 2. Yet, there is no common interest between $u_1$ and $u_2$. 
Similarly, items can be classified into multiple clusters based on their characteristics. To fully exploit the valuable information from neighboring nodes, it is necessary to jointly consider the multiple interests of each user and the common interest among users for capturing accurate and comprehensive representations of users and items, particularly when performing high-order graph convolution operations.

Motivated by the above consideration, we propose a novel recommendation model termed Cluster-based Graph Collaborative Filtering (ClusterGCF). This model performs high-order graph convolution operations on cluster-specific graphs, which are constructed by capturing the multiple interests of users and identifying the common interests among them. 
To construct the cluster-specific graphs, we introduce an unsupervised and optimizable soft node clustering approach that classifies all the user and item nodes into multiple clusters by leveraging their embeddings and the graph topology information. 
Based on the soft node clustering results and the topology of the user-item interaction graph, we assign the nodes with probabilities for different clusters to construct the cluster-specific graphs.
Since the embeddings of nodes are propagated within cluster-specific graphs during the graph convolution process, our model can filter out the negative information while capturing more valuable information from high-order neighboring nodes. Besides, the proposed ClusterGCF maintains the uniqueness of node embeddings while stacking more graph convolution layers because it can benefit from higher-order neighboring nodes assigned with different probabilities in cluster-specific graphs. As a result, the over-smoothing problem could be alleviated. Our model's effectiveness was validated through extensive experimentation on four publicly available datasets. Our ClusterGCF results demonstrate superior performance compared to the state-of-the-art GCN-based recommendation methods. This indicates that our model can benefit from higher-order neighbors by classifying nodes into multiple clusters. Furthermore, the results also verified the effectiveness of our proposed soft node clustering approach. We released the codes and involved parameter settings to facilitate others to repeat this work~\footnote{https://github.com/zhao254014/ClusterGCF.}.

In summary, the main contributions of this work are as follows:
\begin{itemize}
    \item We emphasize that GCN-based recommendation methods overlook the users' multiple interests when aggregating information from high-order neighboring nodes. Inspired by this, we propose a ClusterGCF model that performs high-order graph convolution over cluster-specific graphs. 
	
    \item We develop a soft node clustering approach, which softly classifies the user and item nodes into multiple clusters in an unsupervised and optimizable way. Cluster-specific graphs are constructed based on the clustering results and the topology of the original user-item interaction graph.

    \item We evaluate the ClusterGCF on four benchmark datasets through empirical studies. Results show that ClusterGCF can gain better user and item representations than SOTA GCN-based recommendation models.
\end{itemize}
\section{related work}
In this section, we provide a concise overview of recent developments in the fields of collaborative filtering and GCN-based recommendation models. These methods are closely relevant to our work. 

\subsection{Collaborative Filtering}
Collaborative Filtering (CF) techniques~\cite{Koren2009MF,rendle2009bpr,wang2019kdd,he2017neural,wang2019ngcf,He@LightGCN,wang2022TOIS,Han@MetaMMRS} are widely used in modern recommendation systems and play an essential role in suggesting items to users. 
Their core idea is to learn users' and items' representations ($aka.$ embedding vectors) by exploiting the interaction behavior between them. Among them,
Matrix Factorization (MF)~\cite{Koren2009MF} is a typical method that learns user and item embeddings by reconstructing the user-item interaction matrix. Due to its powerful capabilities, deep learning is widely used in recommendations to learn user and item embeddings~\cite{cheng2016wide,xue2017deep,Li2019CNR}. For example, Wide \& Deep~\cite{cheng2016wide} adopts deep neural networks to learn dense low-dimensional embedding for sparse features; DMF~\cite{xue2017deep} learns a common low-dimensional space for user and item representations using a deep structure learning architecture that combines explicit ratings and implicit feedback. To improve the recommendation performance, the Deep learning technique is further employed to capture fine-grained user preferences. For example, NAIS~\cite{He2018Nais} employs an attention network to estimate the importance of interacted items for user preference; MAML~\cite{liu2018MAML} introduces a multimodal attentive approach to capture user's diverse preferences.
After obtaining the embedding vectors for both the user and the item, a preferred score is calculated between them using the dot product~\cite{rendle2009bpr,Koren2009MF}. Due to the intricate nature of user-item interactions, deep learning techniques are also employed as the interaction prediction function in recommendation systems~\cite{he2017neural}.


Despite great success, the above-mentioned methods are still unable to obtain the optimal embeddings for recommendation. The reason is that the CF signals are only implicitly captured, so the transitivity property of behavior similarity could be captured~\cite{wang2019ngcf}. A common solution is to explicitly leverage the graph structure to assist the user and item representation learning. In the following, we will focus on discussing GCN-based recommendation methods.

\subsection{GCN-based Recommendation Models}
Early recommendation approaches used random walks on graphs to leverage the high-order proximity between users and items, inferring indirect user preference~\cite{BW2015Christoffel,fouss2007random}. Some recent approaches leverage the bipartite graph structure of user-item interactions to enhance these relationships~\cite{yu2018walkranker}. Furthermore, they probe further to uncover collaborative relations, such as similarities between users and items~\cite{chen2019cse}. 
The efficacy of these methods is closely tied to the quality of interactions sampled through the random walks, as they rely heavily on these samples for model training.
In recent years, Graph Convolutional Networks (GCNs) have garnered significant attention in the realm of recommendation systems. Their rise in popularity can be attributed to their adeptness at learning node representations from non-Euclidean structures~\cite{berg2019gcmc,wang2019ngcf,wang2019kdd,wang2020DGCF,huang2021point,He@LightGCN,Liu2021IMP_GCN,Wang2021TOIS,lin2022NCL,Cheng2023MBR,Wei2023LightGT,Liu2023TOIS}. For example, 
NGCF~\cite{wang2019ngcf} propagates embeddings on the user-item interaction graph to harness the benefits of the high-order proximity. Drawing inspiration from studies focused on simplifying GCNs~\cite{pmlr-v97-wu19e}, researchers also critically scrutinize the intricacies of GCN-based recommendation models. Notably, He et al.~\cite{He@LightGCN} observed that the two commonly adopted design elements—feature transformation and nonlinear activation—did not contribute positively to the final performance.  Consequently, they introduced LightGCN, a model that omits these two components and yields a marked improvement in recommendation performance. To achieve more efficient recommendations, the UltraGCN~\cite{mao2021ultragcn} model further simplifies graph convolutional networks by omitting embedding propagation layers. Additionally, it incorporates higher-order relationships through user-user graphs, leading to better performance and shorter running time compared to LightGCN. To further improve the neural graph collaborative filtering,
NCL~\cite{lin2022NCL} explicitly incorporates the structural neighbors and semantic neighbors into contrastive pairs. Since the GCN-based recommendation methods suffer from the over-smoothing problem,
IMP-GCN~\cite{Liu2021IMP_GCN} refines user and item embeddings by conducting high-order graph convolutions within subgraphs. It can benefit from higher-order neighboring nodes by filtering out the noisy information from them. To learn more effective and robust representations, Fan et al.~\cite{fan2022GTN} proposed GTN, which can adaptively determine the reliability of interactions and effectively reduce unreliable interaction behaviors.

In this paper, we argue that the existing GCN-based methods are still insufficient for obtaining valuable information from high-order neighbors since the multiple interests of users are not well considered. To tackle the problem, we proposed an unsupervised and optimizable soft node cluster method that can cluster users and items to build cluster-specific graphs. By performing high-order graph convolution on cluster-specific graphs, more valuable information could be delivered to the target node.


\section{METHODOLOGY}
\subsection {Preliminaries}
Given a set of users $\mathcal{U}$ and a set of items $\mathcal{I}$, and interaction data (e.g., ratings, click-throughs) among users and items, the goal of the recommendation system is to predict a set of items a user did not consume before and would be interested in. 
The interaction matrix, denoted by $\bm{R}$, is a matrix of dimensions $\mathcal{N} \times \mathcal{M}$, where $\mathcal{N}$ and $\mathcal{M}$ are the numbers of users and items, respectively. Each entry $r_{ui}$ represents the interaction of user $u \in \mathcal{U}$ with the item ${i} \in \mathcal{I}$. For binary interactions, $r_{ui} = 1$ if there is an interaction, and 0 otherwise. In our method, every user and item is assigned a unique ID and is represented by an embedding vector.

\textbf{User-item Interaction Graph.} The User-Item Interaction Graph is a bipartite graph that provides a structured representation of the interaction data, enabling algorithms like GCNs to capture intricate patterns. Users, items, and their interactions are represented as a graph, where users and items are nodes, and interactions are edges. The aim is to learn a low-dimensional representation (or embeddings) of the nodes (i.e., users and items) such that it can be used to predict future interactions. In particular, we construct the user-item interaction graph $\mathcal{G}= (\mathcal{W}, \mathcal{E})$ based on the interaction matrix $\bm{R}$, where the node set $\mathcal{W} = \mathcal{U} \cup \mathcal{I}$ consists of user nodes and item nodes, and $\mathcal{E}$ represents the edge set. In addition, we define $\mathcal{N}_u$ as the set of item nodes neighboring user node $u$. Similarly, $\mathcal{N}_i$ denotes the set of user nodes neighboring item node $i$.


\begin{figure}[t]
	\centering
\includegraphics[width=0.9\linewidth]{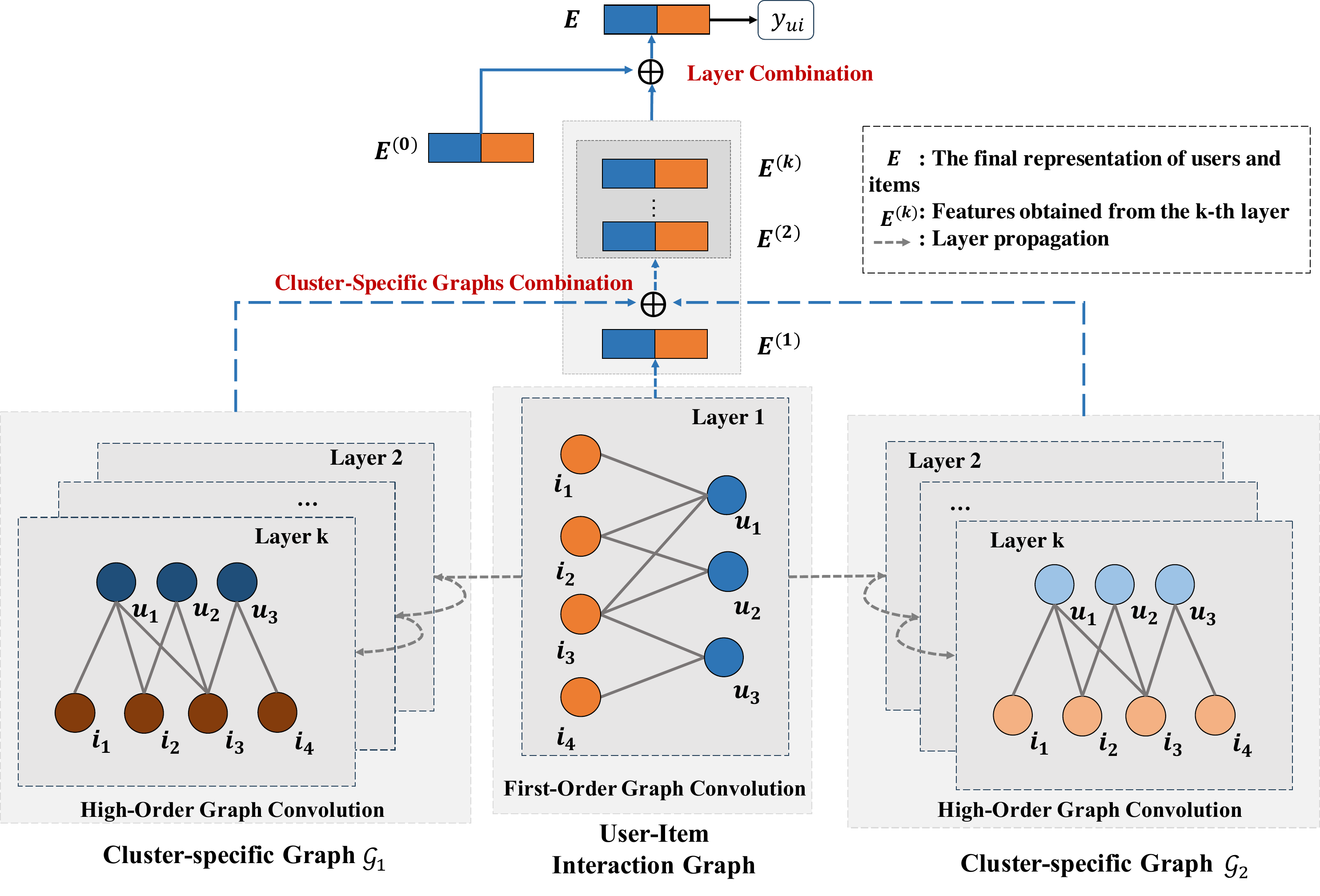}
	\caption{An overview of our ClusterGCF model with two cluster-specific graphs. In ClusterGCF, first-order propagation operates on the original user-item interaction graph, and high-order propagation operates on the cluster-specific graphs $\mathcal{G}_1$ and $\mathcal{G}_2$. The cluster-specific graphs are constructed by assigning probabilities to the nodes of the original user-item interaction graph based on their association with different clusters.}
	\vspace{0pt}
	\label{fig:ClusterGCF_gra}
\end{figure}

\subsection{ClusterGCF Model}
In this section, we present \textbf{Cluster-based Graph Collaborative Filtering} (ClusterGCF) model, along with its training methodology for recommendations. As illustrated in Fig.~\ref{fig:ClusterGCF_gra}, ClusterGCF performs the first-order graph convolution over the original user-item interaction graph~\footnote{Note that the first-order graph convolution is adopted as implemented in IMP-GCN~\cite{Liu2021IMP_GCN}.} and the high-order graph convolution on the cluster-specific graphs to form the final representation of users and items.  To construct cluster-specific graphs, we propose a soft node clustering approach to group users and items into multiple clusters. Next, based on the results of clustering, we assign a probability score to each node in the original user-item interaction graph for each cluster. This allows us to generate a graph that is specific to each cluster and shows the connections between users and items within that cluster.
In the following, we will provide details of the ClusterGCF.

\subsubsection{First-Order Graph Convolution}
Since the direct interactions between users and items provide the most important and reliable information~\cite{Liu2021IMP_GCN}, all first-order neighbors are involved in the graph convolution. In ClusterGCF, the first-order graph convolution is adopted as previous studies~\cite{He@LightGCN,Liu2021IMP_GCN}:
\begin{equation}
\label{first_order_MP}
\begin{aligned}
\bm{e_{u}^{(1)}} =  \sum_{i \in \mathcal{N}_{u}}\frac{1}{\sqrt{ | \mathcal{N}_{u} | }\sqrt{ | \mathcal{N}_{i} |}}\bm{e_{i}^{(0)}}, \\
\bm{e_{i}^{(1)}} =  \sum_{u \in \mathcal{N}_{i}}\frac{1}{\sqrt{ | \mathcal{N}_{i} | }\sqrt{ | \mathcal{N}_{u} |}}\bm{e_{u}^{(0)}}. 
\end{aligned}
\end{equation}
For the target user $u$ and item $i$, $\bm{e_u^{(0)}}$ and $\bm{e_i^{(0)}}$ are ID embeddings, while their first layer embeddings are represented as $\bm{e_u^{(1)}}$  and $\bm{e_i^{(1)}}$. In addition, $\frac{1}{\sqrt{ | \mathcal{N}_u | }\sqrt{ | \mathcal{N}_i |}}$ are symmetric normalization terms.

\subsubsection{High-Order Graph Convolution}
To reduce the noisy information but also assist in aggregating more valuable information from high-order neighboring nodes, we propagate embeddings over multiple cluster-specific graphs. For cluster $c \in \mathcal{N}_c$, where $\mathcal{N}_c$ is the set of clusters, the 2nd order graph convolution on the cluster-specific graph $\mathcal{G}_c$ associated with cluster $c$ is defined as:
\begin{equation}
\begin{aligned}
\bm{e_{u}^{(c, k)}} =  \sum_{i \in \mathcal{N}_{u}}\frac{p_i^c}{\sqrt{ | \mathcal{N}_{u} | }\sqrt{ | \mathcal{N}_{i} |}}\bm{e_{i}^{(1)}}, \\
\bm{e_{i}^{(c, k)}} =  \sum_{u \in \mathcal{N}_{i}}\frac{p_u^c}{\sqrt{ | \mathcal{N}_{i} | }\sqrt{ | \mathcal{N}_{u} |}}\bm{e_{u}^{(1)}},
\end{aligned}
\end{equation}
where $\bm{e_{u}^{(c, k)}}$ and $\bm{e_{i}^{(c, k)}}$ denote the embedding of user $u$ and item $i$ after $k$ layers graph convolution and $k = 2$. $p_u^c$ and $p_i^c$ represent the probabilities assigned to user $u$ and item $i$ for cluster $c$. When $k > 2$, we define the $k-$th order convolution on the cluster-specific graph $\mathcal{G}_c$ as:
\begin{equation}
\begin{aligned}
\bm{e_{u}^{(c, k)}} =  \sum_{i \in \mathcal{N}_{u}}\frac{p_i^c}{\sqrt{ | \mathcal{N}_{u} | }\sqrt{ | \mathcal{N}_{i} |}}\bm{e_{i}^{(c, k-1)}}, \\
\bm{e_{i}^{(c, k)}} =  \sum_{u \in \mathcal{N}_{i}}\frac{p_u^c}{\sqrt{ | \mathcal{N}_{i} | }\sqrt{ | \mathcal{N}_{u} |}}\bm{e_{u}^{(c, k-1)}}.
\end{aligned}
\end{equation}

\textbf{Cluster-Specific Graphs Combination.} With the graph convolution operation, we can obtain the node representations for each layer of cluster-specific graphs. In other words, the messages are passed from the $k-$th layer neighboring nodes based on our constructed cluster-specific graphs. These messages are further combined as the $k-$th layer representations of user $u$ and item $i$ ($k \geq 2$):
\begin{equation}
\begin{aligned}
\bm {e_u^{(k)}} = \sum_{c \in {\mathcal{N}_{c}}}\bm{e_{u}^{(c, k)}},\\
\bm {e_i^{(k)}} = \sum_{c \in {\mathcal{N}_c}}\bm{e_{i}^{(c, k)}},
\end{aligned}
\end{equation}
where $\bm{e_{u}^{(k)}}$ and $\bm{e_{i}^{(k)}}$ denote the features of user $u$ and item $i$ aggregated from the $k$-th layer neighbor nodes, respectively. Previous studies~\cite{He@LightGCN, Liu2021IMP_GCN} have demonstrated the effectiveness of aggregating information from neighboring nodes through embedding summation. Thus, we apply the same method to aggregate the information within and between cluster-specific graphs. Note that we treat the information within different cluster-specific graphs equally because assigning different probabilities to the nodes in the different clusters can capture the varying importance.

\subsubsection{Layer Combination and Prediction.} Due to its effectiveness, we adopt the same layer combination and prediction methodology as LightGCN~\cite{He@LightGCN}. Specifically, after performing $k$ layers of graph convolution, the final embeddings of user $u$ and item $i$ are obtained by combing their embeddings across all layers of the ClusterGCF:
\begin{equation}
\begin{aligned}
\bm{e_{u}}=\sum_{k=0}^{K}\alpha_{k}\bm{e_{u}^{(k)}},\\
\bm{e_{i}}=\sum_{k=0}^{K}\alpha_{k}\bm{e_{i}^{(k)}},
\end{aligned}
\end{equation}
where $\alpha_{k}$ is uniformly set to $1/(K+1)$. 
After learning the representation for both users and items, the preference of a given user $u$ for a target item $i$ is estimated as follows:
\begin{equation}
    \hat{r}_{ui} = \bm{e_{u}}^T\bm{e_{i}}.
\end{equation}

\subsubsection{Soft Node Clustering}
\label{sec:nodeclustering}
\begin{figure}[t]
	\centering
\includegraphics[width=0.7\linewidth]{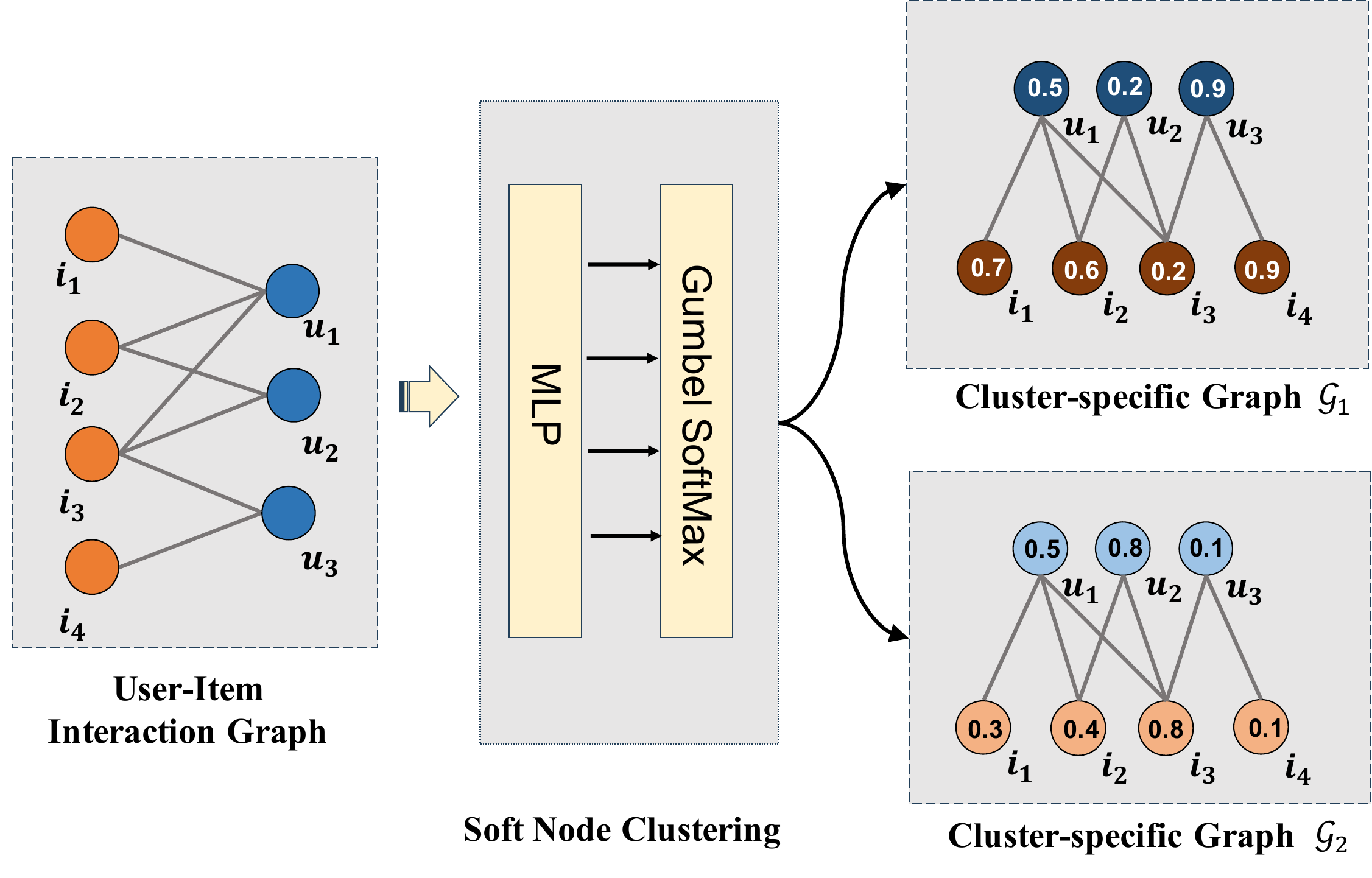}
	\caption{An example of cluster-specific graphs construction. Each user and item node of the user-item interaction graph is classified into two clusters via a soft node clustering method. Specifically, each node in the original graph is allocated to the two clusters based on its soft assignment probabilities. Take user node $u_2$ as an example, it is assigned to cluster 1 with a weight of 0.2 and to cluster 2 with a weight of 0.8. Based on the soft assignment probabilities and the original edges from the user-item interaction graph, two cluster-specific graphs $\mathcal{G}_1$ and $\mathcal{G}_2$ are constructed.}
	\vspace{0pt}
	\label{fig:nodeclustering}
\end{figure}
We introduce a soft node clustering approach to capture the multiple interests of users and identify the common interests among them. 
Since users typically exhibit multiple interests in different aspects of items~\cite{liu2018MAML}, we classify both the user and item nodes into multiple clusters with varying degrees of association instead of assigning them to a single cluster. 
Take a user $u$ and an item $i$ as an example, we start by fusing the graph structure and ID embedding:
\begin{equation}
\begin{aligned}
\label{feature_fusion2}
\bm{F_{u}} = \sigma (\bm{W_{1}}(\bm{e_u^{(0)}} + \bm{e_u^{(1)}}) + \bm{b_{1}}),\\
\bm{F_{i}} = \sigma (\bm{W_{1}}(\bm{e_i^{(0)}} + \bm{e_i^{(1)}}) + \bm{b_{1}}),
\end{aligned}
\end{equation}
where $\bm{F_u}$ and $\bm{F_i}$ are the enhanced features of the user $u$ and item $i$ obtained through feature aggregation. 
$\bm{W_{1}} \in R^{d \times d}$ and  $\bm{b_1} \in R^{1 \times d}$ are the weight matrix and bias vector, respectively. $\sigma$ denotes the activation function for which the LeakyReLU function is adopted~\cite{Andrew2013leaky}.
Subsequently, we cast these enhanced features into prediction vectors using a single-layer neural network for classifying users and items into various clusters:
\begin{gather} 
\begin{aligned}
\begin{split}
\bm {H_{u}} = \sigma(\bm{W_{2}F_{u}} + \bm{b_{2}}),\\
\bm {H_{i}} = \sigma(\bm{W_{2}F_{i}} + \bm{b_{2}}),\\
\end{split}
\end{aligned}
\end{gather}
where $H_u$ and $H_i$ are the prediction vectors for the user and item, respectively. Similarly, $\bm{W_{2}} \in R^{d \times |\mathcal{N}_{c}|}$ and $\bm{b_2} \in R^{1 \times |\mathcal{N}_{c}|}$ are the weight matrices and bias vector. It is worth noting that the dimension of the prediction vector corresponds to the number of cluster-specific graphs, a value determined by a pre-selected hyperparameter.

\textbf{Gumbel-Softmax.}
One of the challenges in clustering methods is the non-differentiability of discrete cluster assignments, which makes gradient-based optimization problematic. To overcome this limitation, we adopt the Gumbel-Softmax technique~\cite{jang2016gumbel_softmax} for soft clustering in our proposed approach. Since the Gumbel-Softmax is differentiable, it can be included directly in end-to-end learning pipelines that involve neural networks. In other words, the clustering approach can be co-optimized with the recommendation model.

The Gumbel-Softmax function provides a smooth, differentiable approximation to the discrete distribution, parameterized by logits and a temperature parameter. In our approach, the last layer of the neural network generates logits for all clusters. These logits are then fed into a Gumbel-Softmax layer to produce soft cluster assignments for each user or item. For example, when given logits $\bm{H_u}$, the Gumbel-Softmax function outputs a probability distribution $\bm{P_{u}}$ over multiple clusters. Mathematically, it is defined as follows:
\begin{equation}
\bm{P_{u}} = \text{Softmax}\left(\frac{\bm{H_u} + g}{\tau}\right),
\end{equation}
where $z$ are the logits, $g$ is the Gumbel noise, and 
$\tau$ is the temperature. The Gumbel noise can be generated through the Gumbel distribution, or more commonly, from uniform noise $U$ using the transformation $-\log(-\log(U))$. The temperature parameter 
$\tau$ allows fine control over the approximation. Lower 
$\tau$ values approximate a hard, one-hot encoded assignment, while higher $\tau$ values produce a more uniform, exploratory distribution.
The Softmax function transforms the modified logits into a probability distribution:
\begin{equation}
\text{Softmax}(x_i) = \frac{e^{x_i}}{\sum_{j} e^{x_j}}.
\end{equation}
Our clustering approach can be trained end-to-end using gradient-based optimization techniques, owing to the Gumbel-Softmax function. It classifies users and items into different clusters in an unsupervised manner, without requiring ground-truth labels.

\textbf{Cluster-Specific Graph Construction.}
After employing soft clustering techniques, specifically using a Gumbel-Softmax trick, we reconstruct the original user-item interaction graph into a series of cluster-specific graphs. 
Unlike conventional methods that may use thresholding to determine hard cluster assignments, our approach retains the granularity of soft assignments. 
Each node in the original graph is allocated to all clusters based on its soft assignment probabilities.
A cluster-specific graph is defined as follows:
\begin{itemize}
\item \textbf{Nodes:} The user and item nodes from the original graph are weighted by their soft assignment probabilities to the given cluster.

\item \textbf{Edges:} The original edges of the user-item interaction graph are retained, thus preserving the innate relationships between users and items.
\end{itemize}

The differentiation between these cluster-specific graphs lies in the weights (soft assignment probabilities) associated with each node.  
As shown in Fig.~\ref{fig:nodeclustering}, each user node $u_{x}$ or item node $i_{x}$ is softly classified into two clusters with different probabilities. The relationship between the probabilities in the two clusters can be described as follows:
\begin{equation}
\begin{aligned}
\label{feature_fusion1}
p_1(u_{x}) + p_2(u_{x}) = 1,\\
p_1(i_{x}) + p_2(i_{x}) = 1,
\end{aligned}
\end{equation}
where $p_1(\cdot)$ and $p_2(\cdot)$ denote the probabilities assigned to this node for these two cluster-specific graphs, respectively.

\subsection{Model Training}
\subsubsection{Optimization.}
In this work, our goal is to recommend a list of items to the target users by computing their preference score. For the optimization, we adopt the pairwise learning method, which is similar to the approach employed in previous works~\cite{Liu2021IMP_GCN}. To train our ClusterGCF model, we construct a triplet of $\{u, i^+, i^-\}$, where $i^+$ represents a positive item and $i^-$ denotes a negative item. The objective function for ClusterGCF is defined as:
\begin{equation}
\mathop{\arg\min} \sum_{(\mathbf{u}, \mathbf{i}^+,\mathbf{i}^-)\in{\mathcal{O}}} -\ln\phi(\hat{r}_{ui^+} - \hat{r}_{ui^-}) + \lambda\left\|\Theta\right\|^2_2
\end{equation}
where $\mathcal{O}$ represent the training set, defined as $\mathcal{O}=\{(u, i^+, i^-)|(u,i^+)\in\mathcal{R^+}, (u,i^-) \in\mathcal{R^-}\}$. Here, $\mathcal{R^+}$ indicates the observed interactions in the training dataset, while $\mathcal{R^-}$ denotes a set of sampled unobserved interactions. Additionally, $\lambda$ represents the regularization weight, and $\Theta$ stands for the model parameters. We employ $L_2$ regularization to mitigate overfitting. To optimize our model and update its parameters, we utilize the mini-batch Adam technique, as described in~\cite{kingma2014adam}. 

Note that the user and item embeddings serve as inputs for both the node clustering and the recommendation system. We optimize these embeddings, which serve different purposes but are intricately linked within our optimization framework. Our approach ensures that the clustering process is continuously informed by the recommendation task, leading to clusters that are more relevant to the recommendation objective. Conversely, the embeddings benefit from the structure provided by the clusters, resulting in more robust and informative representations of users and items.

\subsubsection{Matrix Form Propagation Rules.}
In order to improve the efficiency of model training, we employ a matrix form propagation rule~\cite{wang2019ngcf} to refine the user and item representations while performing graph convolution operations. Let $\bm{E}^{(0)}$ be the representation matrix for user and item IDs,
the first-layer embedding propagation of ClusterGCF is described as follows:
\begin{equation}
\bm{E^{(1)}}=\mathcal{L} \times \bm{E^{(0)}},
\end{equation}
where $\bm{E}^{(1)}$ is the obtained representation matrix of users and items in the first graph convolution layer and $\mathcal{L}$ is the Laplacian matrix for the user-item interaction graph. 

For the higher-order graph convolution operation, the higher-order neighbor embeddings are first propagated on cluster-specific graphs. For the cluster-specific graph $\mathcal{G}_c$, $\bm{E_c^{(k)}}$ is defined as the obtained representation of users and items in the $k$-th cluster-specific graph convolution layer ($k\geq 2$). This process can be formulated as follows:
\begin{align}
  \bm E_{c}^{k}=\mathcal{L} \times (\bm{{P_c}} \odot \bm{E_{c}^{k-1}}),
\end{align}
where $\bm{P_c}$ denotes the probability vector of all nodes assigned to the cluster-specific graph $\mathcal{G}_c$. 
Then, we aggregate the embedding in the $k$-th layer from all cluster-specific graphs to formulate user and item representations, denoted as $\bm{E^{(k)}}$, in the $k$-th graph convolution layer of ClusterGCF:
\begin{equation}
\bm E^{k}=\sum_{c\in \mathcal{N}_{c}} E_{c}^{k}.
\end{equation}
Finally, we combine the embeddings from all layers to obtain the final representations:
\begin{equation}
\bm{E} = \alpha_{0}\bm{E^{(0)}} + \alpha_{1}\bm{E^{(1)}} + \cdots + \alpha_{K}\bm{E^{(K)}}.
\end{equation}

\section{Experiments}

To evaluate the effectiveness of our proposed ClusterGCF, we conducted extensive experiments on four public datasets. In particular, we primarily answer the following questions.

\textbf{RQ1:} Does ClusterGCF outperform state-of-the-art CF baselines in the top-$n$ recommendation task?
 
\textbf{RQ2:} How do the number of graph convolution layers impact ClusterGCF?

\textbf{RQ3:} How do the number of clusters and the starting layer of Cluster-based Graph Convolution affect the performance and model Convergence of ClusterGCF? 

\textbf{RQ4:} How do the temperature coefficient and node types affect the cluster-specific graph construction?

\textbf{RQ5:} How do the node representations benefit from the high-order connectivities in cluster-specific graphs? 


\subsection{Experimental setup}
\subsubsection{Data Description.}
\begin{table}[t]
	\centering
	\caption{ Statistics of the Experimental Datasets.}
	\label{tab:data}
	\begin{tabular}{|c|c|c|c|c|}
		\hline 
		Dataset&\#user&\#item&\#interactions&sparsity \\ \hline \hline 
		Office & 4,874 & 2,406 & 52,957 &99.55\% \\ \hline 
		Kindle Store & 14,356 & 15,885 & 367,477 & 99.83\% \\ \hline 
		Gowalla & 29,858 & 40,981 & 1,027,370 & 99.92\% \\ \hline 
        Yelp2018 & 31,668 & 38,048 &1,561,406 & 99.87\%\\
		\hline 
	\end{tabular}
	\vspace{0pt}
\end{table}
To assess ClusterGCF model, extensive experiments are conducted on the following four datasets: Amazon review datasets~\footnote{http://jmcauley.ucsd.edu/data/amazon.}, Gowalla and Yelp2018~\footnote{https://www.yelp.com/dataset.}. 

\begin{itemize}

\item \textbf{Amazon Review Datasets.} This dataset stems from the extensive product review system of Amazon~\cite{liu2018MAML}. It includes user-product interactions, reviews, and metadata. The interaction information is utilized in our experiments. Two product categories from the 5-core version of this dataset are selected for evaluation: Office and Kindle Store. This indicates that both users and items have a minimum of 5 interactions.

\item \textbf{Gowalla.} This dataset is collected from Gowalla~\footnote{https://www.gowalla.com.}, where users share their locations by checking-in~\cite{wang2019ngcf}. It typically contains several millions of check-ins spread across users and locations. Each check-in also associates a user with a particular location, which is used to model user preferences in our experiments.

\item \textbf{Yelp2018.} This dataset is a subset of the Yelp challenge of 2018. It contains a variety of information mainly focused on businesses, reviews, and users. Businesses, such as restaurants and services, are treated as the items in our experiments~\cite{wang2019ngcf}.
\end{itemize}

To ensure the data quality, we filtered users and items with limited interactions following the general setting in the recommendation~\cite{he2017neural,liu2018MAML}. Specifically, for the Kindle Store, Gowalla, and Yelp2018 datasets, we implemented a 10-core setting, retaining only users and items with a minimum of 10 interactions. The details and statistics of these four datasets are shown in Table~\ref{tab:data}. From the table, the datasets vary in size, which allows us to analyze our method and its competitors' performance under different scenarios.
For each dataset, we randomly split each user's interactions into training and testing sets at an 80:20 ratio. We select 10\% of interactions in the training set randomly to create the validation set for tuning hyper-parameters. The pairwise learning strategy is adopted for parameter optimization. Specifically, the observed user-item interactions were considered positive instances in the training set. For each positive instance, we paired it with a random sampling negative instance that the user has not interacted with yet. For a fair comparison, we consistently adopt the negative sampling strategy across our methods and competitors.

\subsubsection{Evaluation metrics.}
In this work, we adopted three metrics for the evaluation: \textbf{Recall@$\bm{K}$}~\cite{Liu2021IMP_GCN}, \textbf{HR@$\bm{K}$ (Hit Ratio)}~\cite{he2017neural}, and \textbf{NDCG@$\bm{K}$ (Normalized Discounted Cumulative Gain)}~\cite{he2015trirank}.
For each metric, we assess the performance of our method and all competitors based on the top 20 items in the ranking list, with $K = 20$. 

\subsubsection{Compared Baselines.}
To verify the effectiveness of ClusterGCF, we compared it with several competitive methods, mainly considering representative GCN-based models like GCMC, NGCF, LightGCN and UltraGCN, as well as those learning with different advanced techniques, such as disentangled learning ($i.e.$, DGCF), contrastive learning ($i.e.$, NCL), or those also considering noise information like GTN and IMP-GCN.


\begin{itemize}
\item \textbf{NeuMF~\cite{he2017neural}}: 
This stat-of-the-art neural collaborative filtering method captures the non-linear interactions between user and item features by introducing multiple hidden layers over the combined user and item embeddings.

\item \textbf{GCMC~\cite{berg2019gcmc}}: This method uses a single graph convolutional layer based on the GCN technique~\cite{kipf2017gcn} to take advantage of direct connections between users and items. 

\item \textbf{NGCF~\cite{wang2019ngcf}}: This is the very first GCN-based recommendation model that incorporates high-order connectivity of user-item interactions. It encodes the collaborative signal from high-order neighbors by performing embedding propagation on the user-item bipartite graph.

\item \textbf{LightGCN~\cite{He@LightGCN}}: This model can be considered a simplified version of NGCF~\cite{wang2019ngcf}, where certain components such as the feature transformation and the non-linear activation module are removed. The model structure's simplicity can help avoid overfitting, especially when data is not massive.

\item \textbf{UltraGCN~\cite{mao2021ultragcn}}: This model simplifies LightGCN~\cite{He@LightGCN} by eliminating infinite layers of message passing, leading to more efficient recommendations. Specifically, it approximates multiple graph convolution layers and incorporates different node relationships flexibly through a constraint loss.

\item \textbf{DGCF~\cite{wang2020DGCF}}: This GCN-based method refines the interaction graphs considering the various user intents. Meanwhile, it encourages the independence of representations with different intents by adopting the disentangled representation technique.

\item \textbf{IMP-GCN~\cite{Liu2021IMP_GCN}}: This method conducts high-order graph convolution on sub-graphs, constructed based on users with shared interests and associated items. By differentiating high-order neighbors, it could alleviate the over-smoothing problem in GCN-based methods.

\item \textbf{NCL~\cite{lin2022NCL}}: This is a model-agnostic contrastive learning framework. It seeks to capture the correlation between a node and its prototype to improve the neural graph collaborative filtering. 

\item \textbf{GTN~\cite{fan2022GTN}}: This GNN-based model captures the adaptive reliability of interactions between users and items, drawing inspiration from the concept of trend filtering and graph trend filtering.
\end{itemize}


\subsubsection{Parameter settings.}
We used Tensorflow~\footnote{https://www.tensorflow.org.} to implement ClusterGCF. For all models, the embedding size was fixed at 64 and initialized using Xavier~\cite{Xavier2010xavier}. For optimization, we used Adam~\cite{kingma2014adam} with a default learning rate of 0.001. The mini-batch size is 1024 on Office and Kindle Store. In contrast, it is increased to 2048 on Gowalla and Yelp208 to accelerate the model training. The $L_{2}$ regularization coefficient $\lambda$ is searched in $\{1e^{-6},1e^{-5},\cdots,1e^{-1}\}$. The embedding propagation layer is searched in the range of $\{1,2,\cdots,8\}$. The GCMC method uses a single convolutional layer as its original setting. For ClusterGCF and IMP-GCN, the cluster number is searched in the range of $\{2,3,4\}$. Besides, the temperature coefficient is tuned amongst $\{1e^{-2},1e^{-1},\cdots,1e^{2}\}$. The early stopping and validation strategies used are identical to those employed in IMP-GCN, ensuring that we carefully tuned the key parameters.

\subsection{Performance Comparison (RQ1)}
\subsubsection{Overall comparison}
\begin{table*}[t]
	\vspace{0pt}
	\caption{Performance comparison between our ClusterGCF model and the baselines on four datasets. Please note that the values are presented as percentages, with the '\%' symbol omitted.} 
	\centering
	\resizebox{1.0\textwidth}{!}{
		\begin{tabular}{|l|ccc|ccc|ccc|ccc|} \hline
			Datasets	& \multicolumn{3}{c|}{Office} & \multicolumn{3}{c|}{Kindle Store} & \multicolumn{3}{c|}{Gowalla} & \multicolumn{3}{c|}{Yelp2018} 
            \\ \cline{2-4}  \cline{5-7} \cline{8-10} \cline{11-13}
			Metrics	& Recall	&HR &	NDCG	& Recall	&HR & NDCG	& Recall  &HR  &	NDCG	& Recall    &HR &	NDCG	\\ \hline \hline 
			NeuMF	& 6.92	& 21.04 &	6.29  & 6.83   & 20.19 & 4.98 &	12.95	& 45.65 &	11.18  &	4.70	& 31.12 &	3.72  \\ \hline 
			GCMC	&	7.15 & 21.76	&	6.85  & 7.73   & 20.88 & 5.22	&	14.07	& 48.86 &	11.78  &	4.82	& 31.89 &	3.87  \\  
            NGCF	&	12.57 & 32.44 &	7.42  & 9.41   & 25.39 & 5.68	&	15.70	& 52.78 &	13.27  &	5.81	& 36.64 &	4.75  \\  
			LightGCN &	13.19 & 33.20	&   7.83  & 9.94   & 26.53 & 6.08	&	18.09	& 58.83 &	15.42 &	6.33	& 38.52 &	5.12  \\  

    UltraGCN	&	13.52	& 33.87 &   8.13	& 10.70  & 28.49 & 6.63  &	18.60	& 59.60 &	15.72  	&	6.70	& 40.51 &	5.47  \\ \hline
    
			DGCF	&	13.27 & 33.23	&   7.89  & 10.05   & 27.05 & 6.15	&	18.24	& 58.83 &	15.51  	&	6.37	& 38.83 &	5.08  \\  
			IMP-GCN	&	13.48 &	33.84 &   8.11	& 10.77   & 28.61 & 6.75  &	18.62	& 59.77 &	15.76  	&	6.66	& 40.43 &	5.45  \\
           
            NCL	&	13.40	& 33.35 &   8.08	& 10.65   & 28.22 & 6.57  &	18.57	& 59.43 &	15.66  	&	6.63	& 40.12  &	5.38  \\
			GTN	   &	\underline{13.54}	& \underline{34.14} &   \underline{8.16}	& \underline{10.78}   & \underline{28.76} & \underline{6.78}  &	\underline{18.65}	& \underline{59.97}  &	\underline{15.83}  	& \underline{6.75}  & \underline{40.75}  &	\underline{5.54}	\\ \hline 
			ClusterGCF	&	\textbf{13.97$^*$}	& \textbf{35.02$^*$} & \textbf{8.46$^*$}	&	\textbf{11.16$^*$}	& \textbf{29.57$^*$}	 & \textbf{7.11$^*$} & \textbf{19.07$^*$}	&	\textbf{61.31$^*$}	  & \textbf{16.21$^*$}	& \textbf{7.06$^*$}	&	\textbf{42.36$^*$}	 &	\textbf{5.83$^*$}		\\ 
			Improv.  & 3.18\% & 2.58\%  & 3.68\% & 3.53\% & 2.82\%  & 4.87\% & 2.25\% & 2.23\% & 2.4\% & 4.59\% & 3.95\%  & 5.23\% 
			\\ \hline
	\end{tabular}}
	\label{tab:results}
	\vspace{0pt}
\end{table*}
In this section, we compared the performance of our ClusterGCF with all the adopted competitors. Table~\ref{tab:results} shows the experimental results in terms of Recall@20, HR@20, and NDCG@20 for ClusterGCF and all compared methods over four different datasets. The top two performing results are bolded for clarity. From these results, several key observations arise:

NeuMF is a DNN-based model and does not perform as effectively as the GNN-based models tested in our experiments. This shortfall can be attributed to its lack of explicit consideration for both first- and higher-order connectivity between users and items in representation learning. In contrast, graph-based approaches like GCMC perform better than NeuMF, demonstrating the benefits of leveraging graph structure. However, GCMC underperforms NGCF, because it is limited to utilizing only first-order neighbor nodes. As the very first GCN-based collaborative filtering model, NGCF brought into play the high-order connectivity of user-item interactions through GCN techniques. This innovation in embedding high-order connectivities offers a performance boost. LightGCN consistently outperforms both GCMC and NGCF. It simplifies the NGCF by eliminating redundant operations and reducing computational complexity. This showcases the strength of GCN and emphasizes the significance of integrating high-order information into representation learning. 
UltraGCN further simplifies the GCNs. It skips the embedding propagation layers, leading to highly efficient recommendation. As a result, UltraGCN surpasses LightGcn across all datasets.

DGCF yields better performance than LightGCN because it uses a disentangled representation to learn independent and robust user and item embeddings by taking into account the diverse intents of users. Both IMP-GCN, NCL and GTN consistently outperform all the above GCN-based methods over all the datasets. This is largely due to their ability to effectively filter out noisy information from neighboring nodes. 
The enhanced performance of IMP-GCN over LightGCN and DGCF emphasizes the significance of differentiating nodes among high-order neighbors during the graph convolution operation.
On the other hand, GTN achieves promising results by capturing the adaptive reliability of the interaction between users and items, which helps in reducing the negative impact of unreliable information.

Our ClusterGCF method consistently surpasses all baseline models across all datasets in terms of all metrics. It effectively avoids the noise and captures more valuable information from high-order neighbor nodes, alleviating the over-smoothing problem in the GCN-based methods. Moreover, the improvement over IMP-GCN demonstrates the effectiveness of our node clustering method. Its superior performance can be attributed to the following reasons: (1) ClusterGCF considers multiple interests of each user while performing high-order graph convolutions. (2) The proposed soft node clustering approach effectively captures the multiple interests of users, identifies the common interests among them, and can be optimized with the recommendation model in an end-to-end manner.

\subsection{Effect of Graph Convolution Layers (RQ2)}
\begin{figure*}[t]
    \centering
	\hspace{0.0cm}
    {\includegraphics[width=0.32\linewidth]{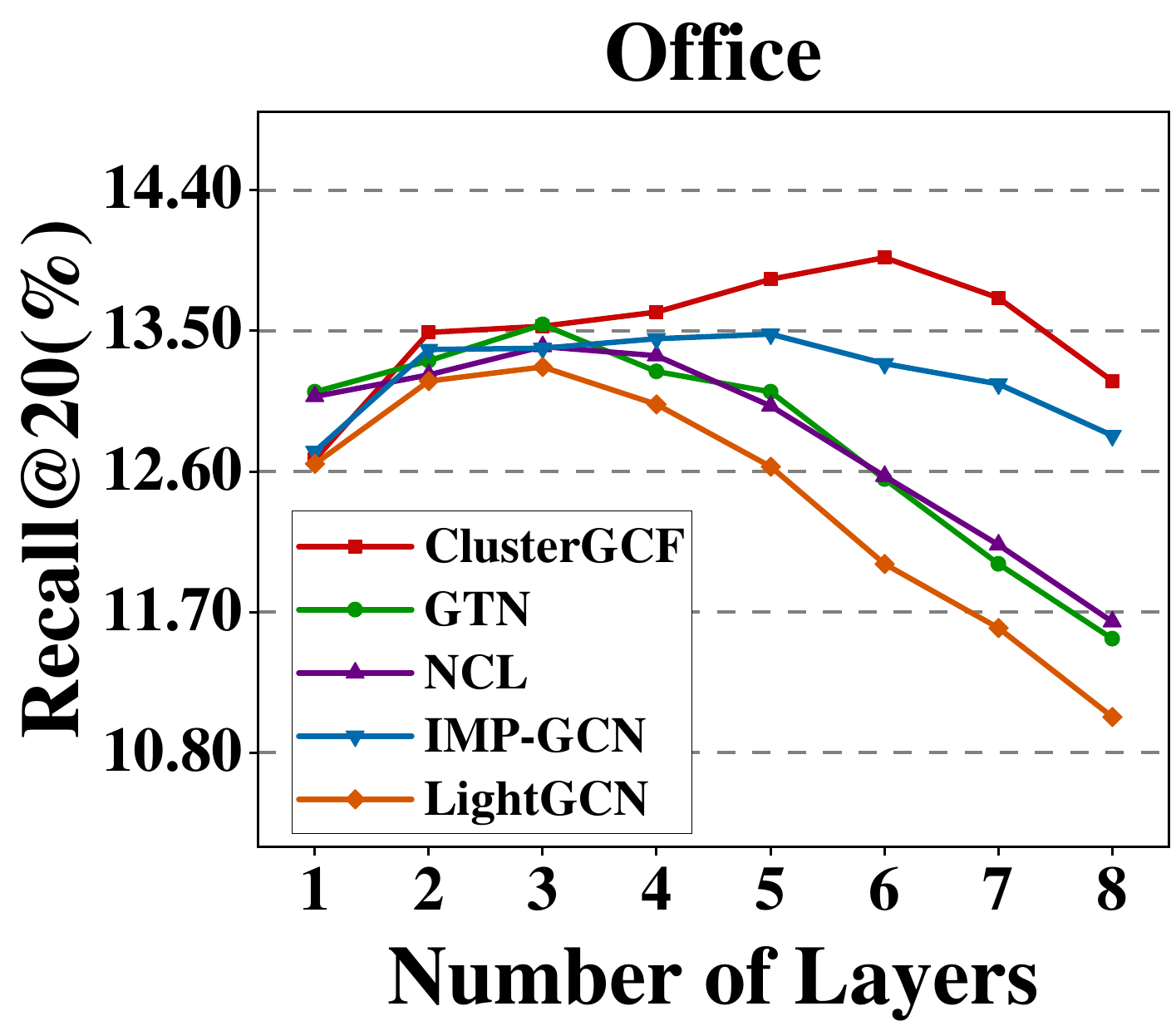}}
    {\includegraphics[width=0.32\linewidth]{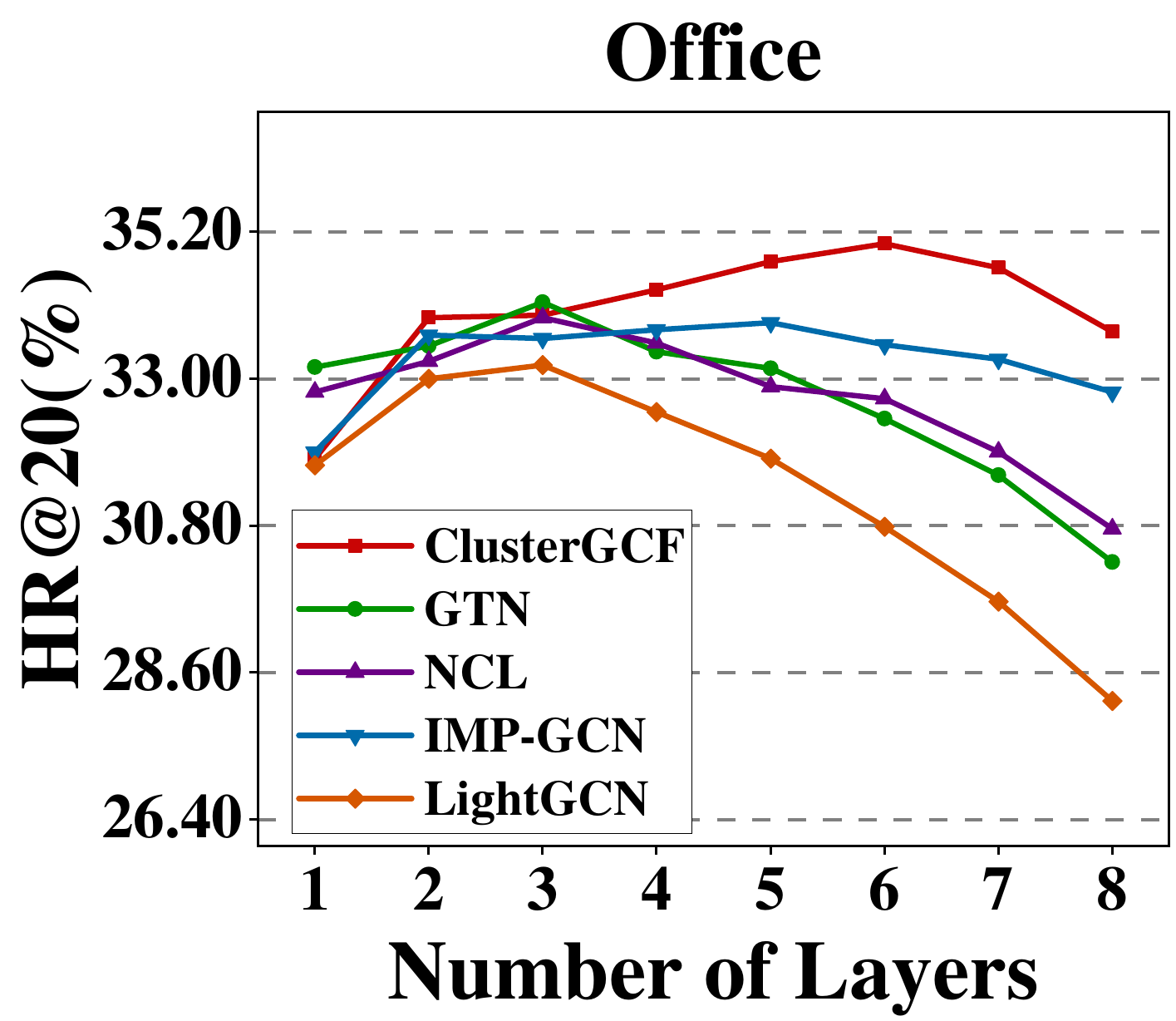}}
    {\includegraphics[width=0.32\linewidth]{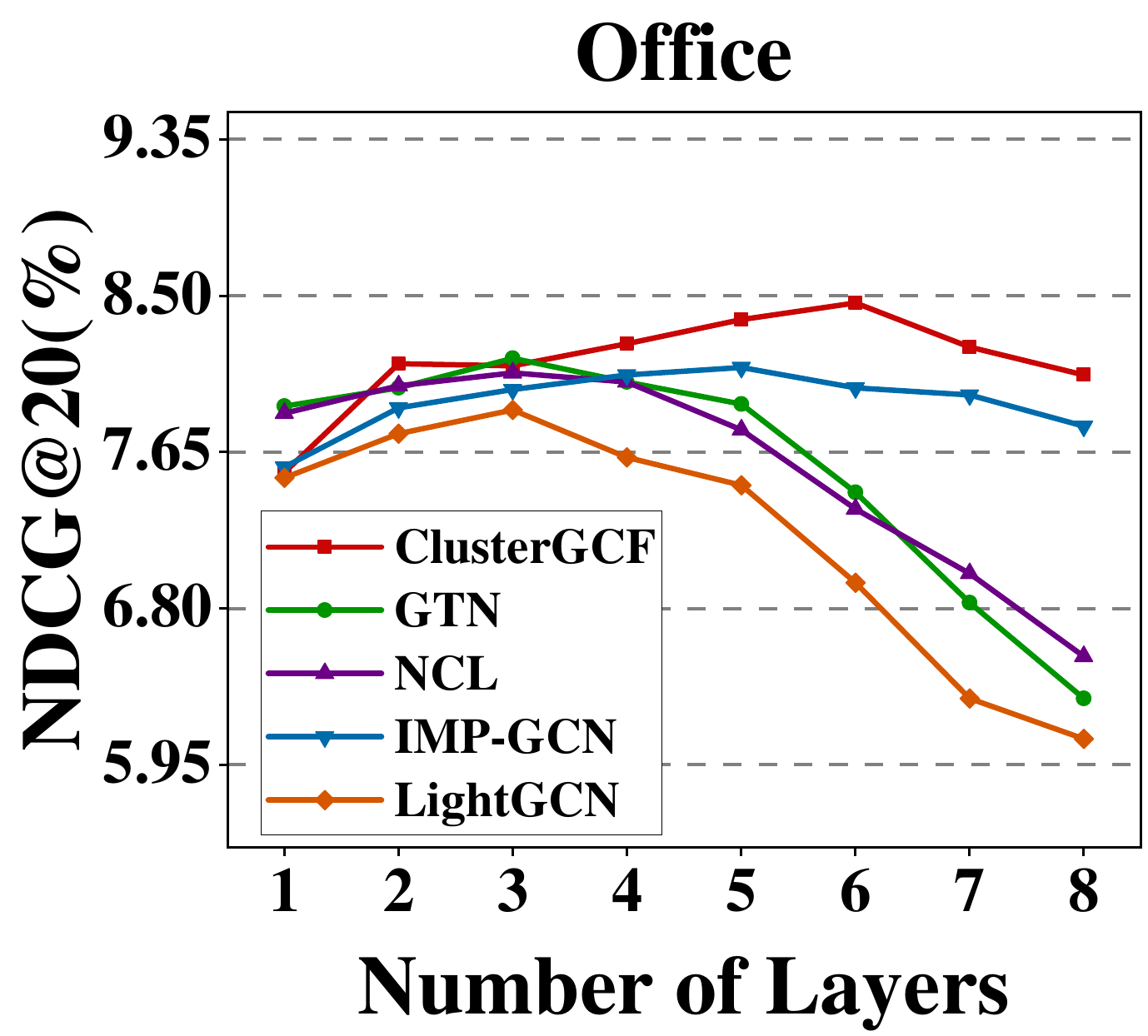}}\\
    {\includegraphics[width=0.32\linewidth]{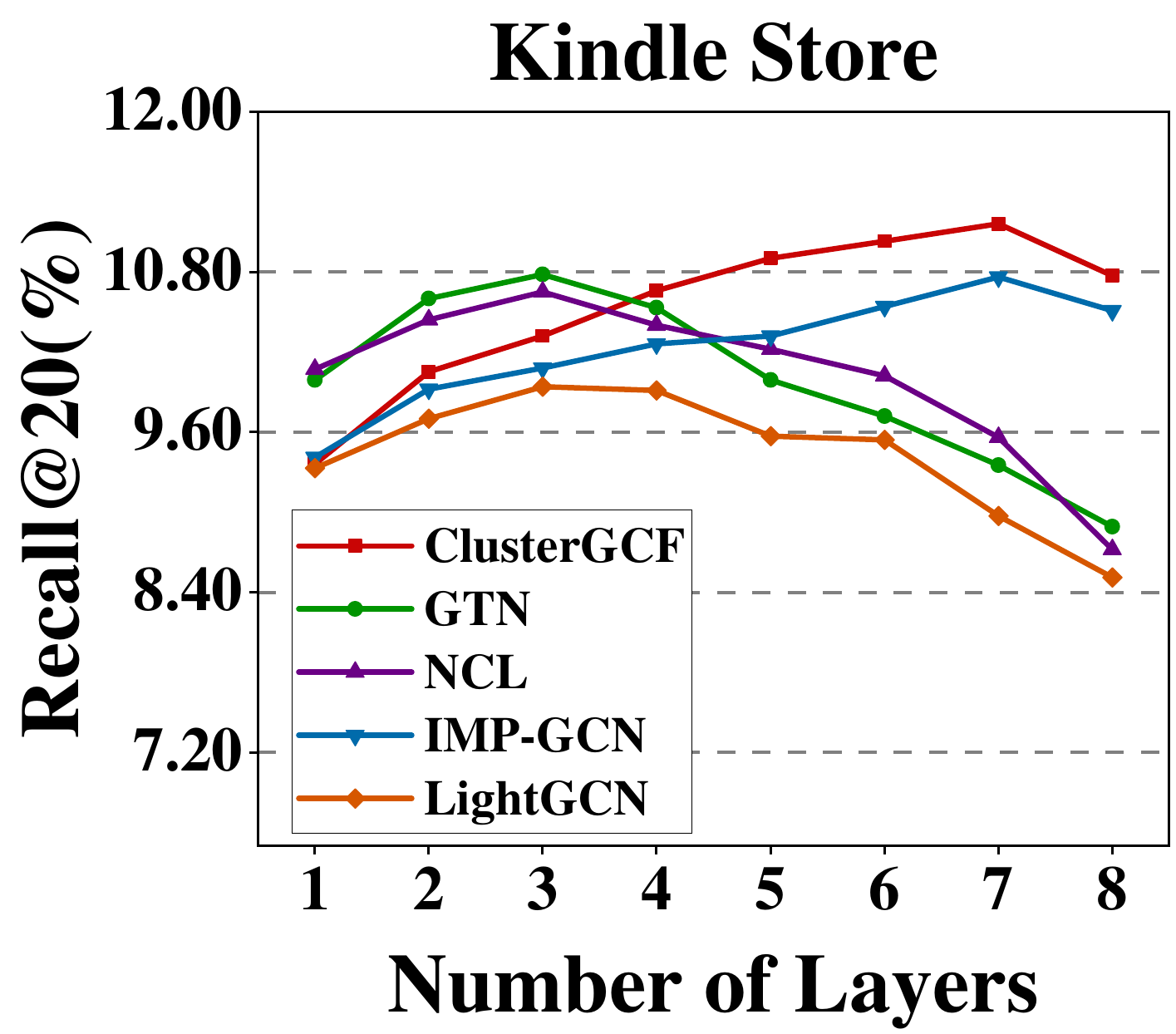}}
    {\includegraphics[width=0.32\linewidth]{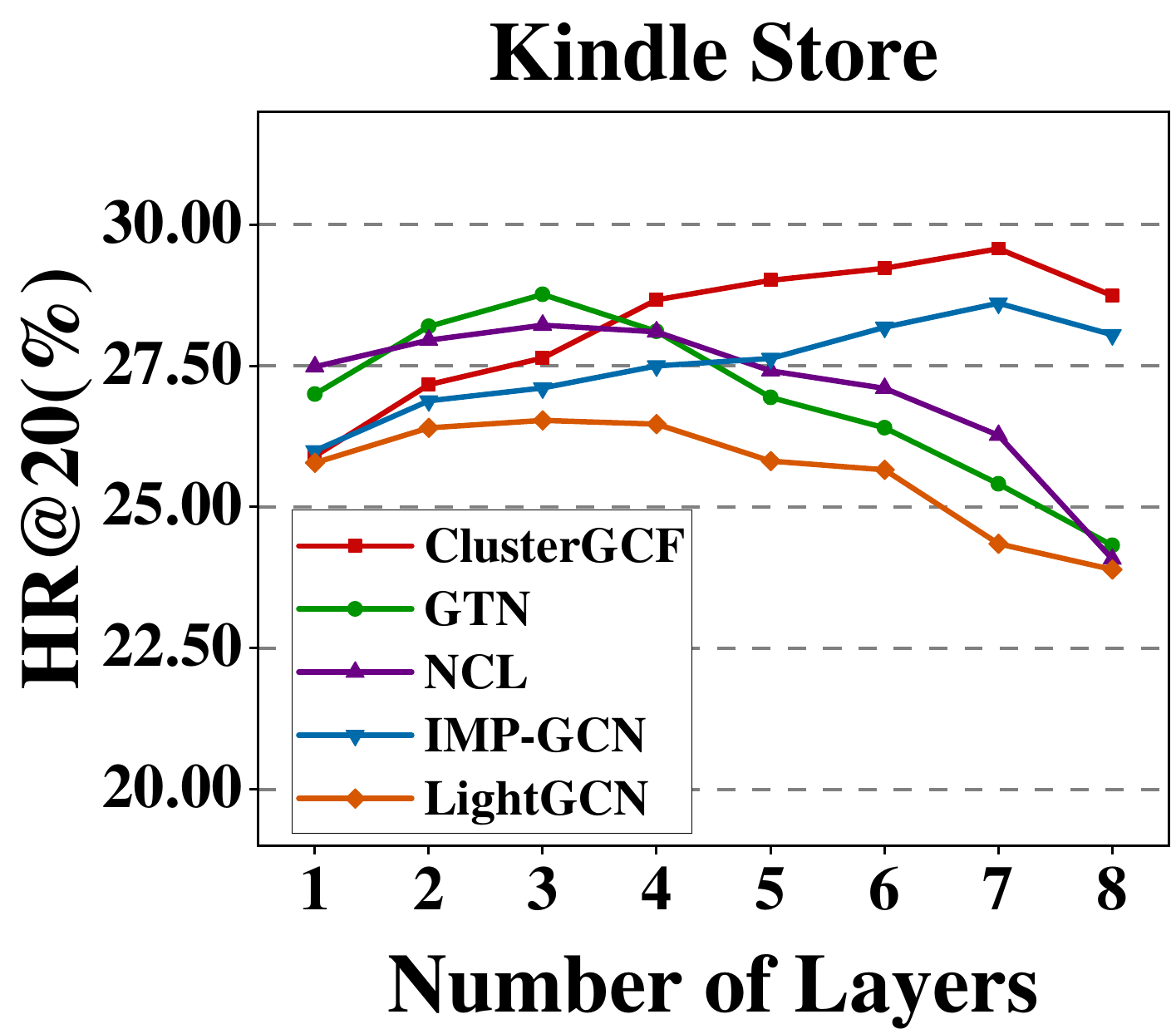}}
    {\includegraphics[width=0.32\linewidth]{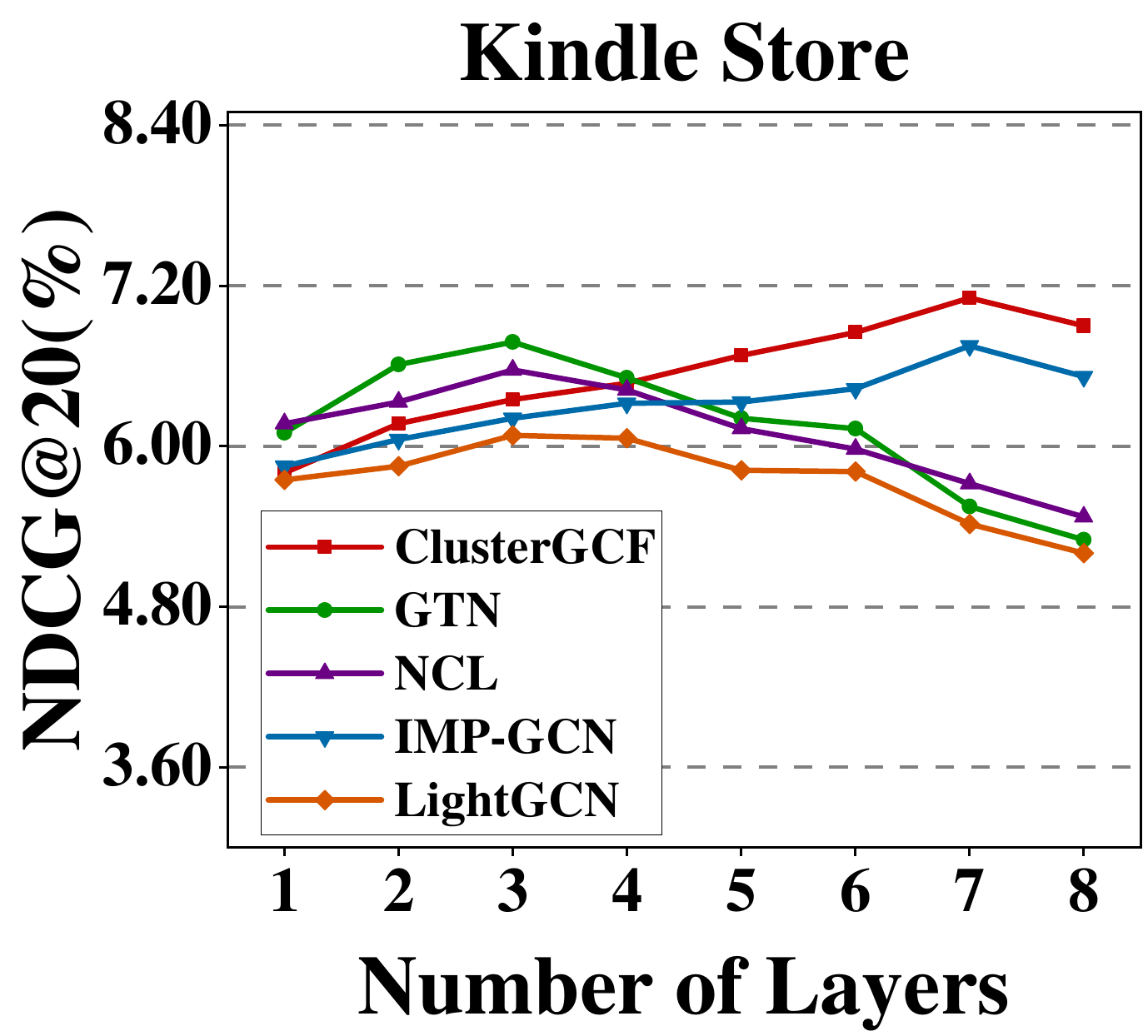}}\\
    {\includegraphics[width=0.32\linewidth]{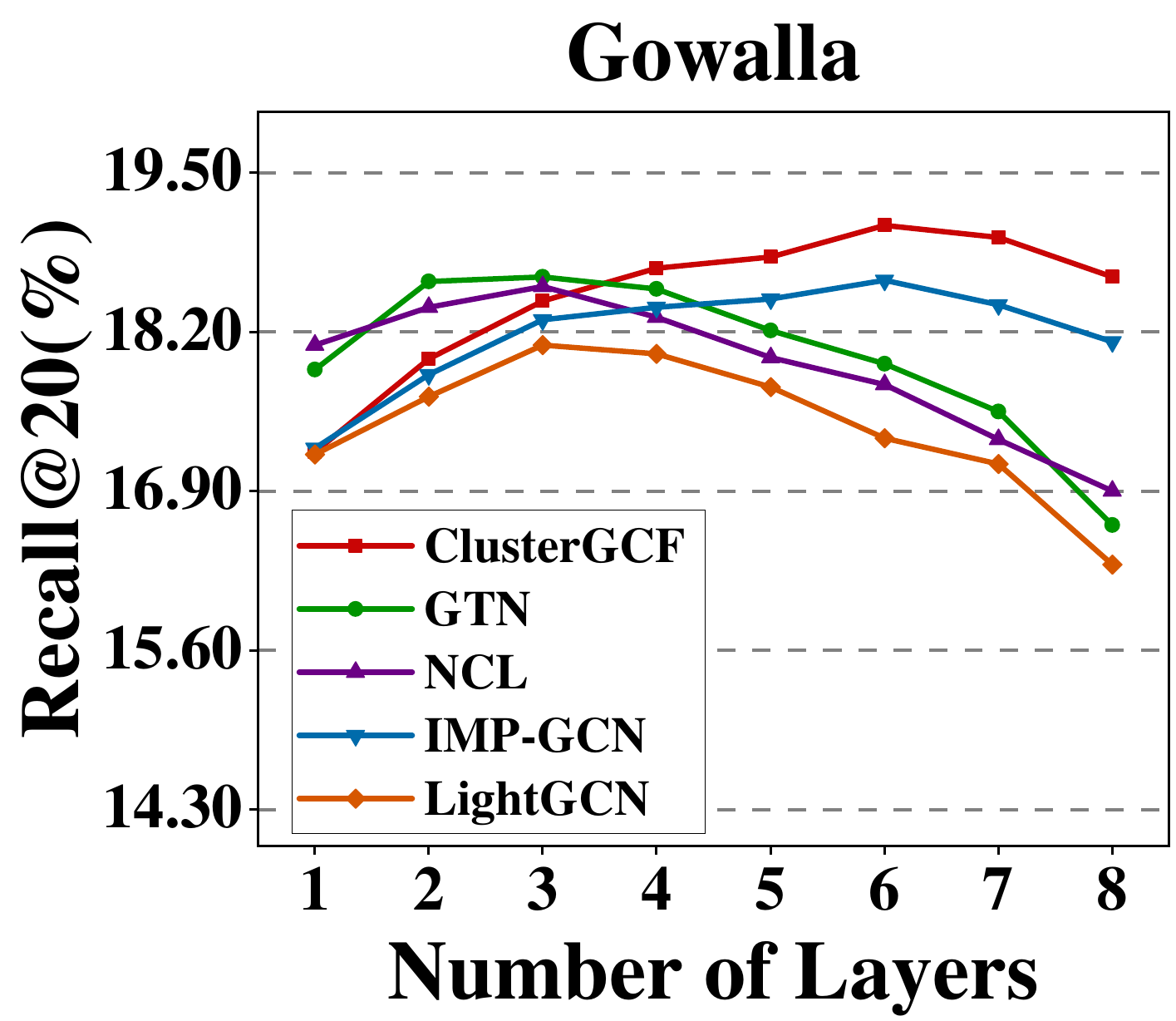}}
    {\includegraphics[width=0.32\linewidth]{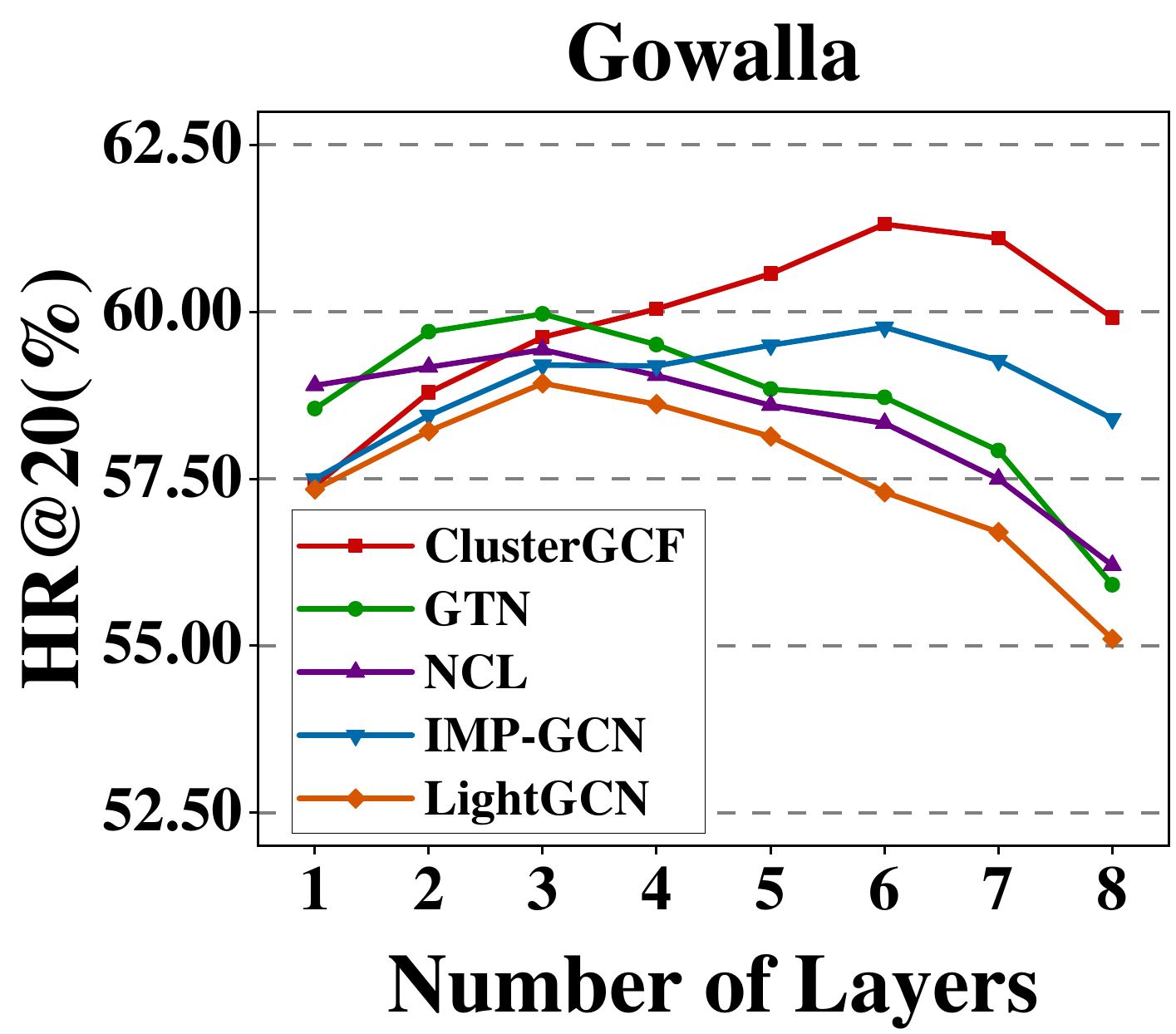}}
    {\includegraphics[width=0.32\linewidth]{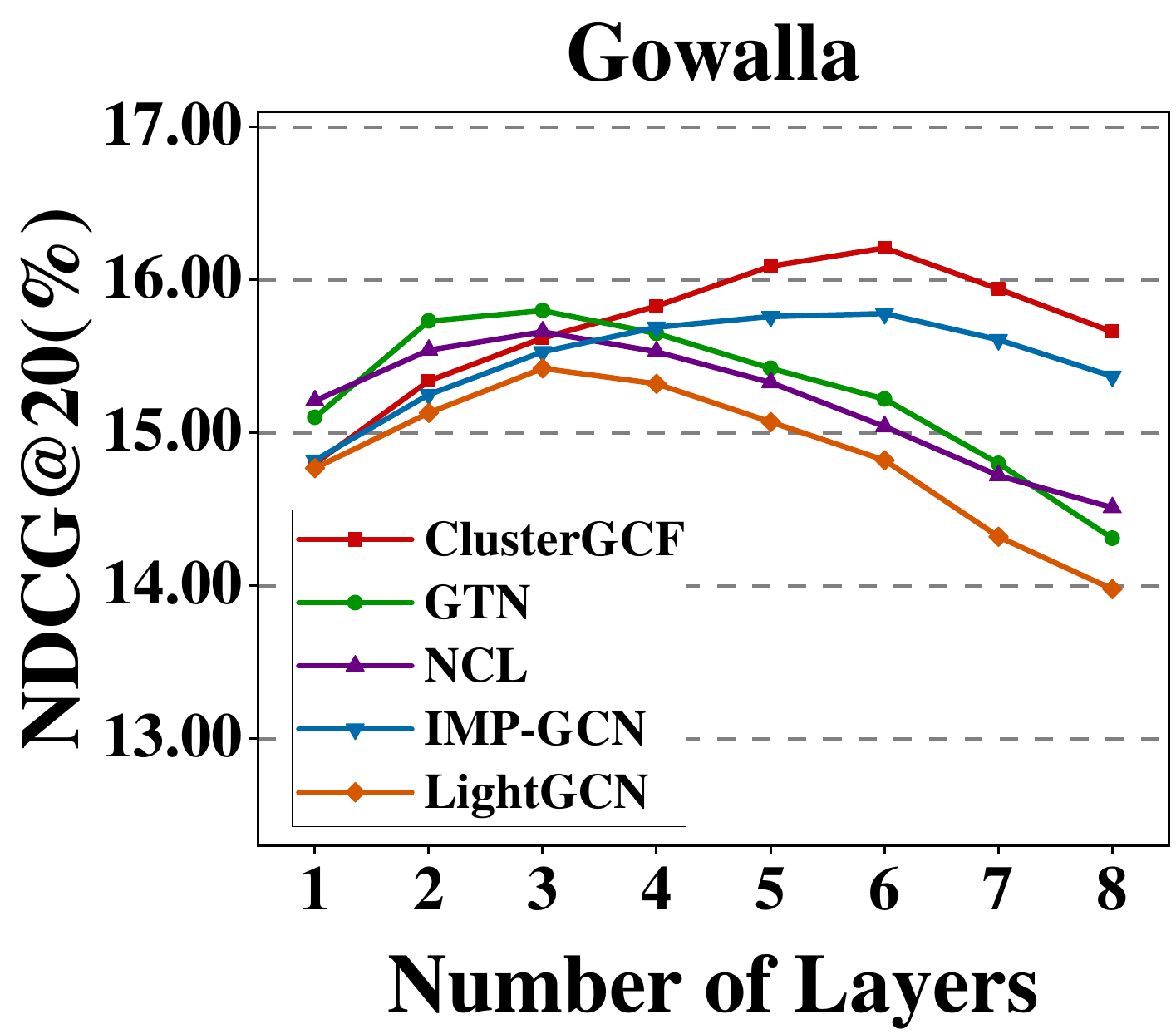}}\\
    {\includegraphics[width=0.32\linewidth]{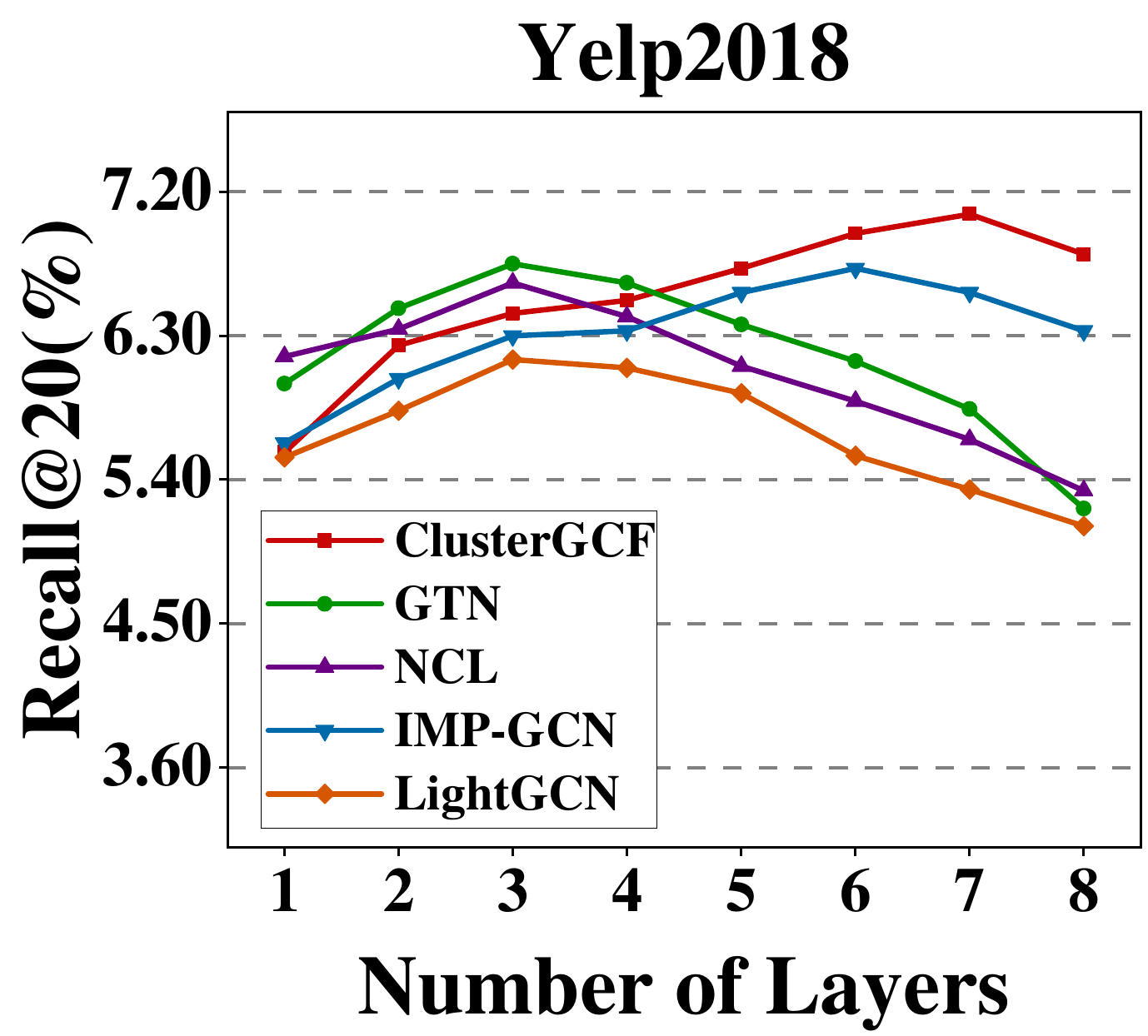}}
    {\includegraphics[width=0.32\linewidth]{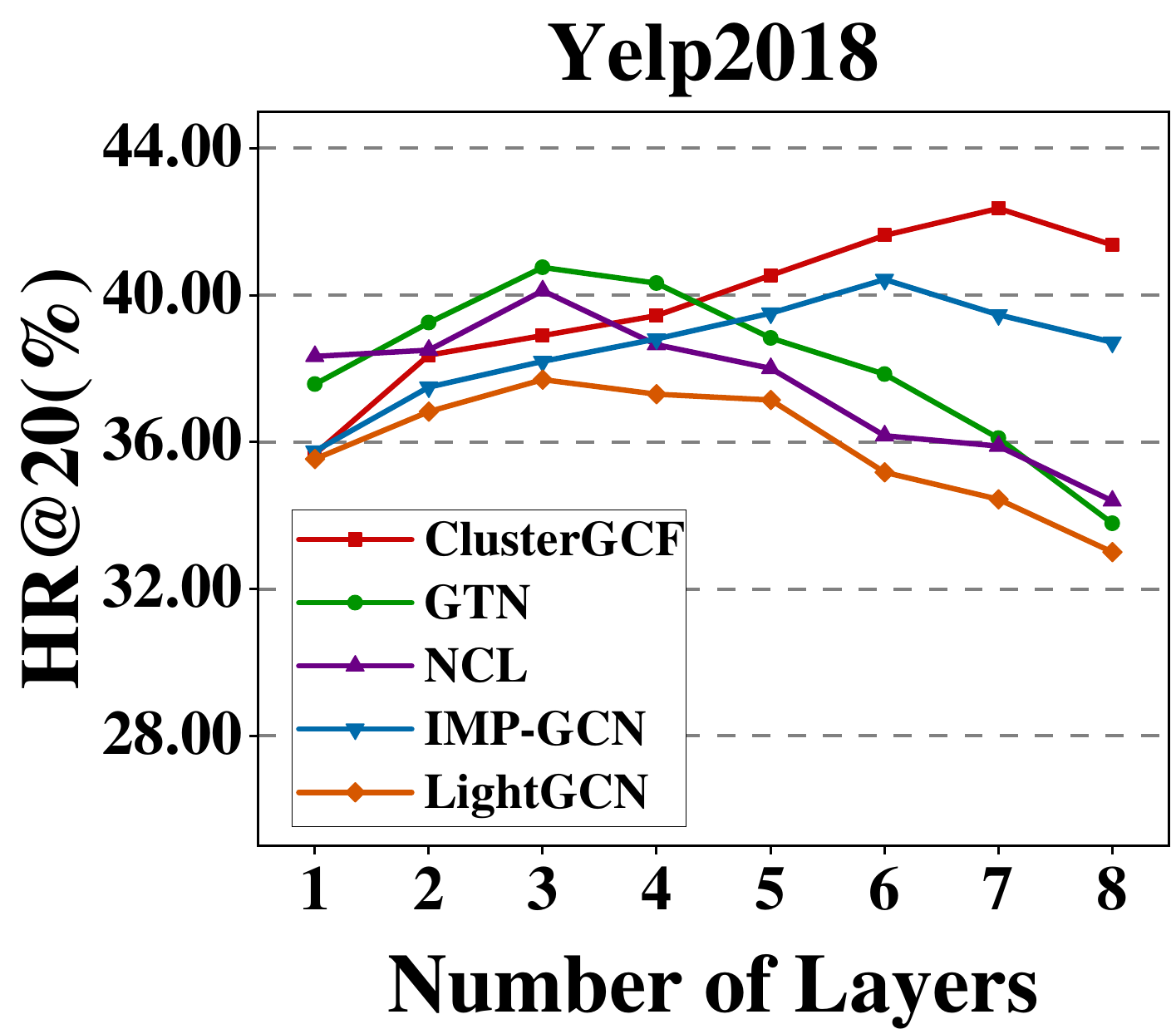}}
    {\includegraphics[width=0.32\linewidth]{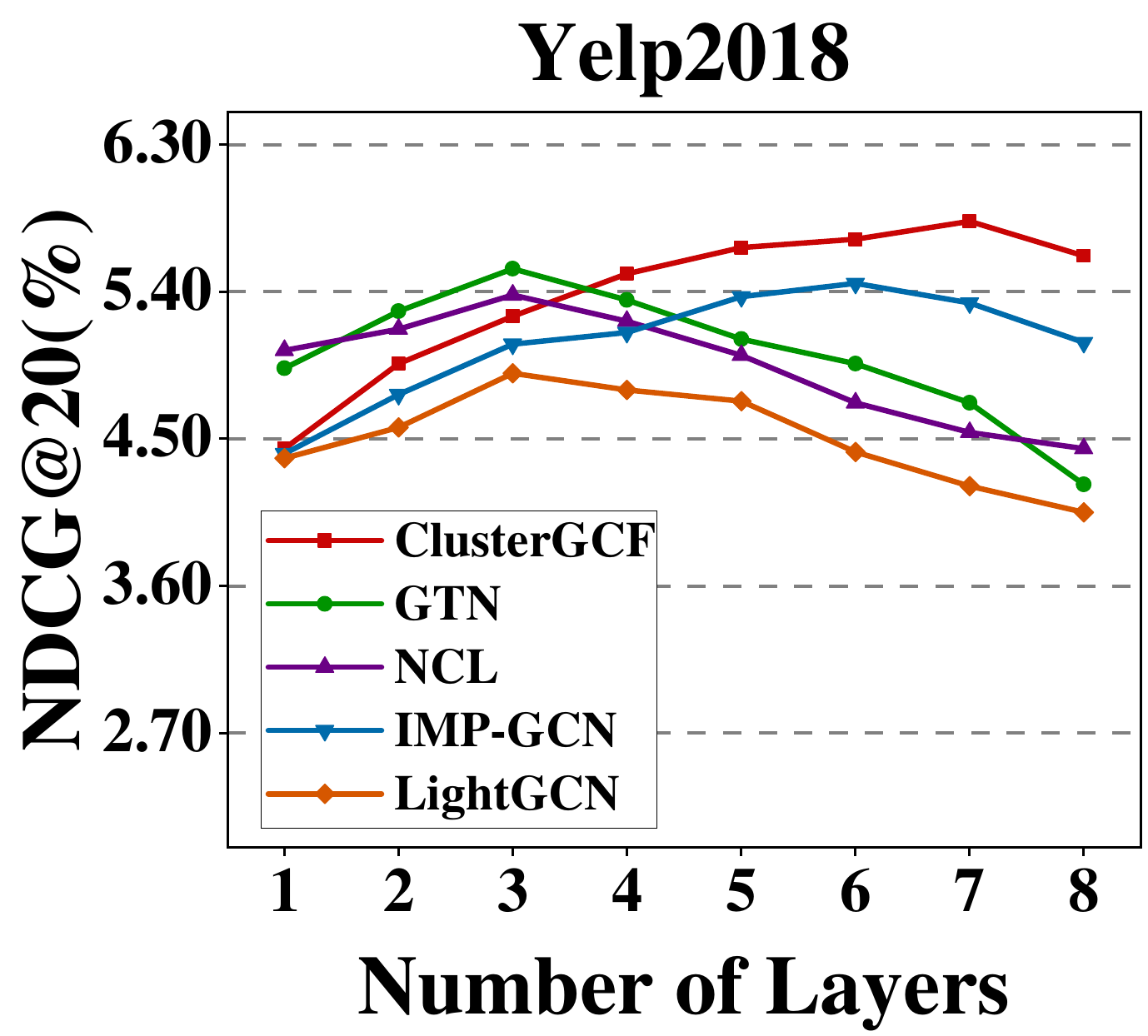}}
	\vspace{0pt}
	\caption{Performance comparison between ClusterGCF and Competitors at different layers on all datasets.}
	\vspace{0pt}
	\label{fig:result-zhe}
\end{figure*}
In this section, we evaluated the performance of our ClusterGCF model by varying the number of layers in the graph convolution process. Our primary goal is to examine whether the ClusterGCF model can mitigate the over-smoothing problem often encountered in GCN-based recommendation models.
To investigate the effectiveness of ClusterGCF, we iteratively increased the number of convolution layers from 1 to 8. We then compared its performance against four GCN-based methods: LightGCN, IMP-GCN, NCL, and GTN. The comparative outcomes are depicted in Fig.~\ref{fig:result-zhe}. In particular, both ClusterGCF and IMP-GCN attained their optimal performance while performing first-order graph convolution across the entire graph. IMP-GCN performs high-order graph convolution within the subgraphs (see Section~\ref{exp:subgraps}) while ClusterGCF operates within the cluster-specific graphs. As a result, both two models adopt the same message-passing strategy as LightGCN in the first-order convolution layer. From the results, we had some interesting observations.

First, all three models (LightGCN, NCL, and GTN) achieve their peak performance when stacking 3 or 4 layers across all datasets. Beyond this, they will encounter dramatic performance degradation with increasing layers. These results indicate that these three models suffer from the over-smoothing problem and inadvertently introduce noise information in the deep structure. Both GTN and NCL achieve better performance than LightGCN at all layers. Moreover, they outperform IMP-GCN when stacking less than 3 or 4 layers. This is because GTN reduces unreliable information by capturing the adaptive reliability of the interactions, and NCL benefits from the semantic and structural neighbors among users/items in representation learning.

Second, IMP-GCN surpasses LightGCN across all datasets when more than 2 or 3 layers are stacked. Significantly, it outperforms GTN and NCL consistently when stacking more than 4 or 5 layers. This suggests that IMP-GCN benefits from a deeper architecture. This resilience can be credited to its effectiveness in filtering out noisy information from high-order neighbors. Additionally, it can also alleviate the over-smoothing problem by performing graph convolution within subgraphs~\cite{Liu2021IMP_GCN}. This demonstrates the significance of distinguishing user nodes in high-order neighbors during the graph convolution operation.

Moreover, the proposed ClusterGCF surpasses all the above-mentioned methods after stacking more than 4 layers across all datasets. This indicates that ClusterGCF can obtain more valuable information from higher-order neighbors. Specifically, ClusterGCF can achieve better performance than IMP-GCN with a deeper structure. For instance, its optimal performance is observed when stacking 6 and 7 layers over Office and Yelp2018, respectively. In contrast, IMP-GCN peaks with 5 and 6 layers on the same datasets. This superiority stems from ClusterGCF distinguishing both user and item nodes in high-order neighbors. Besides, a learnable clustering method is adopted for node classification. It classifies nodes into different clusters in a soft manner. Consequently, ClusterGCF can filter out noise information while capturing more valuable information passing from high-order neighbors.

\subsection{Effect of Cluster-specific Graphs (RQ3)}
\label{exp:subgraps}
\begin{figure*}[t]
    \centering
	\hspace{0.0cm}
	{\includegraphics[width=0.24\linewidth]{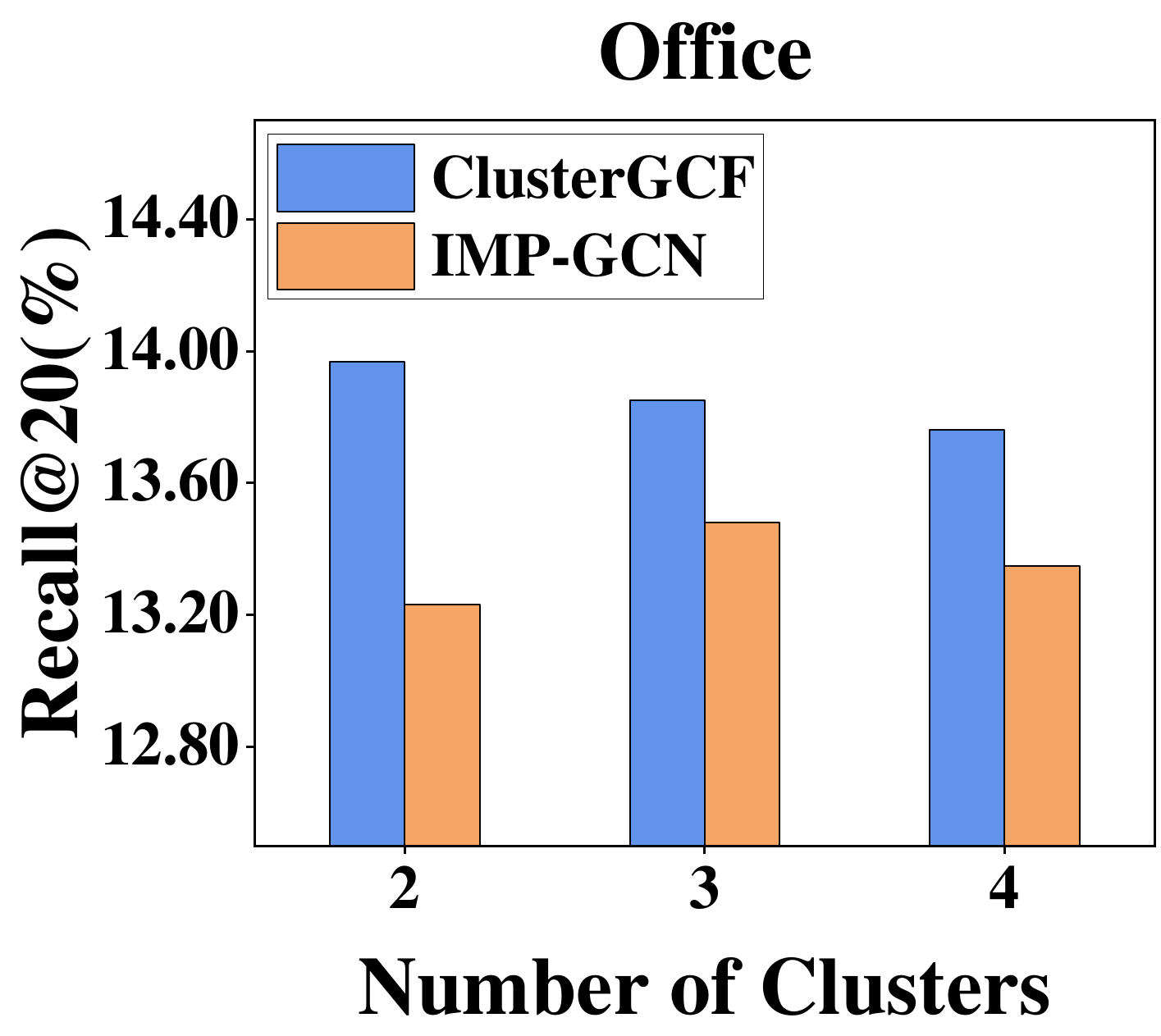}}
    {\includegraphics[width=0.24\linewidth]{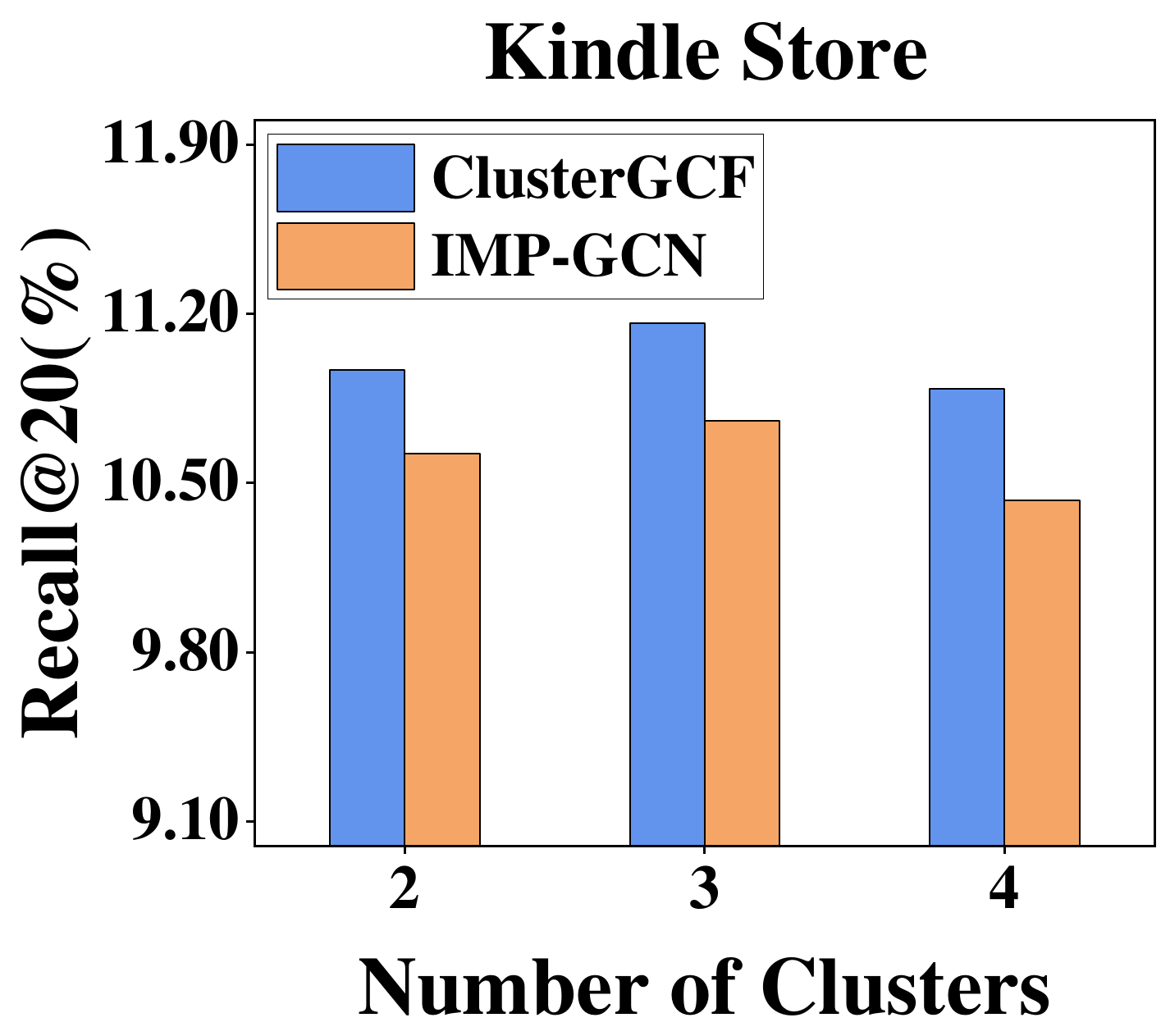}}
    {\includegraphics[width=0.24\linewidth]{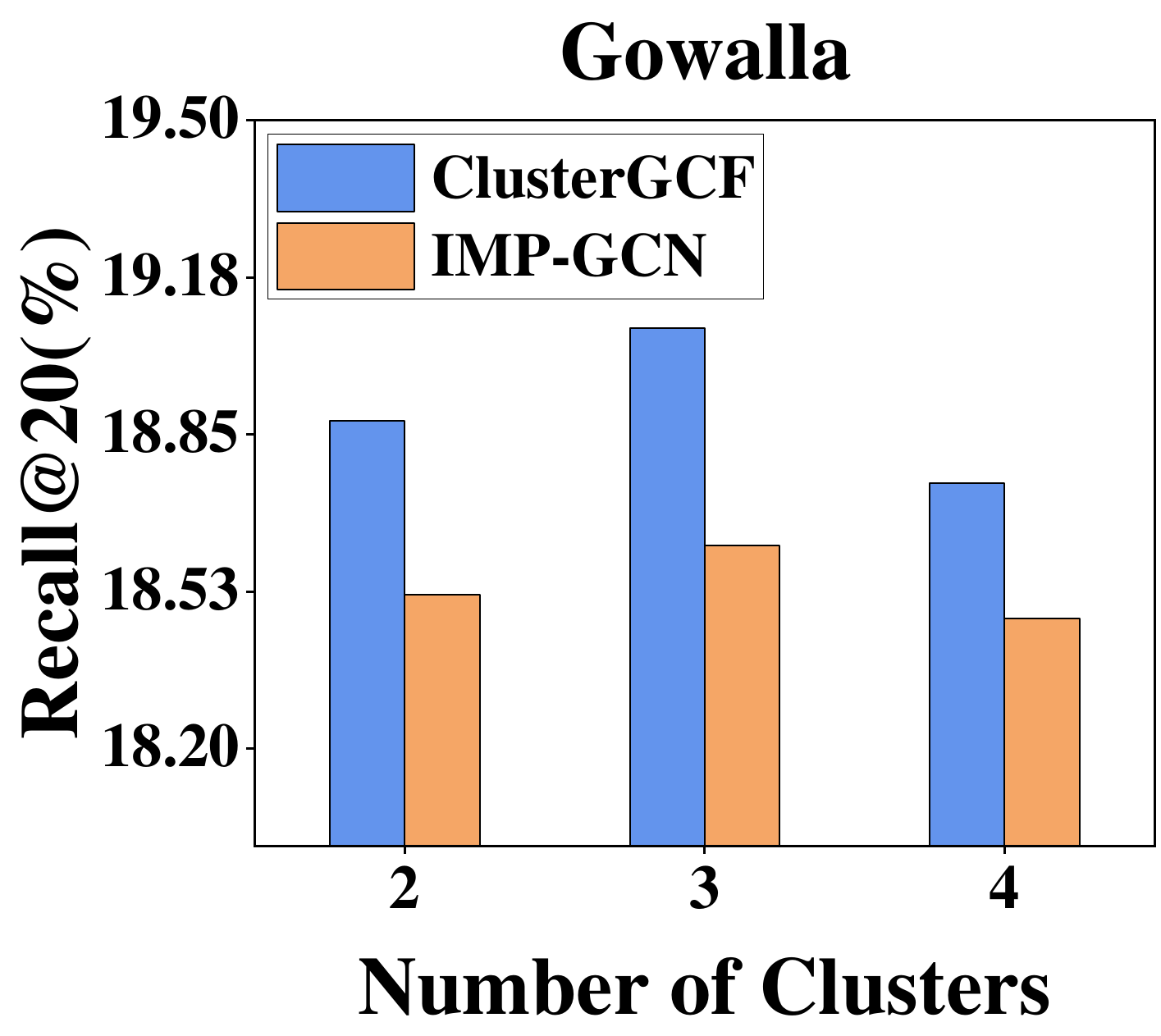}}
    {\includegraphics[width=0.24\linewidth]{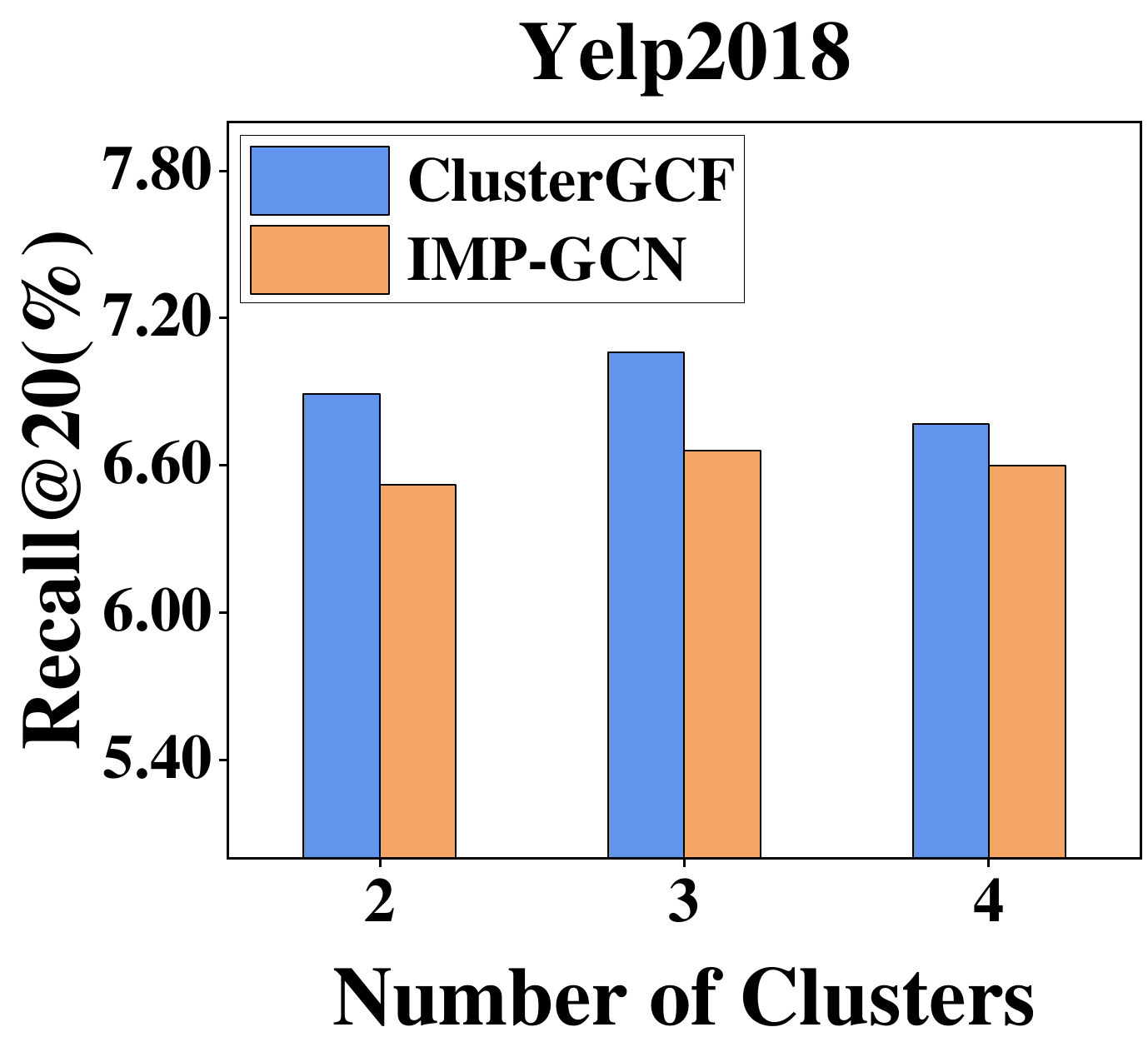}}\\
    {\includegraphics[width=0.24\linewidth]{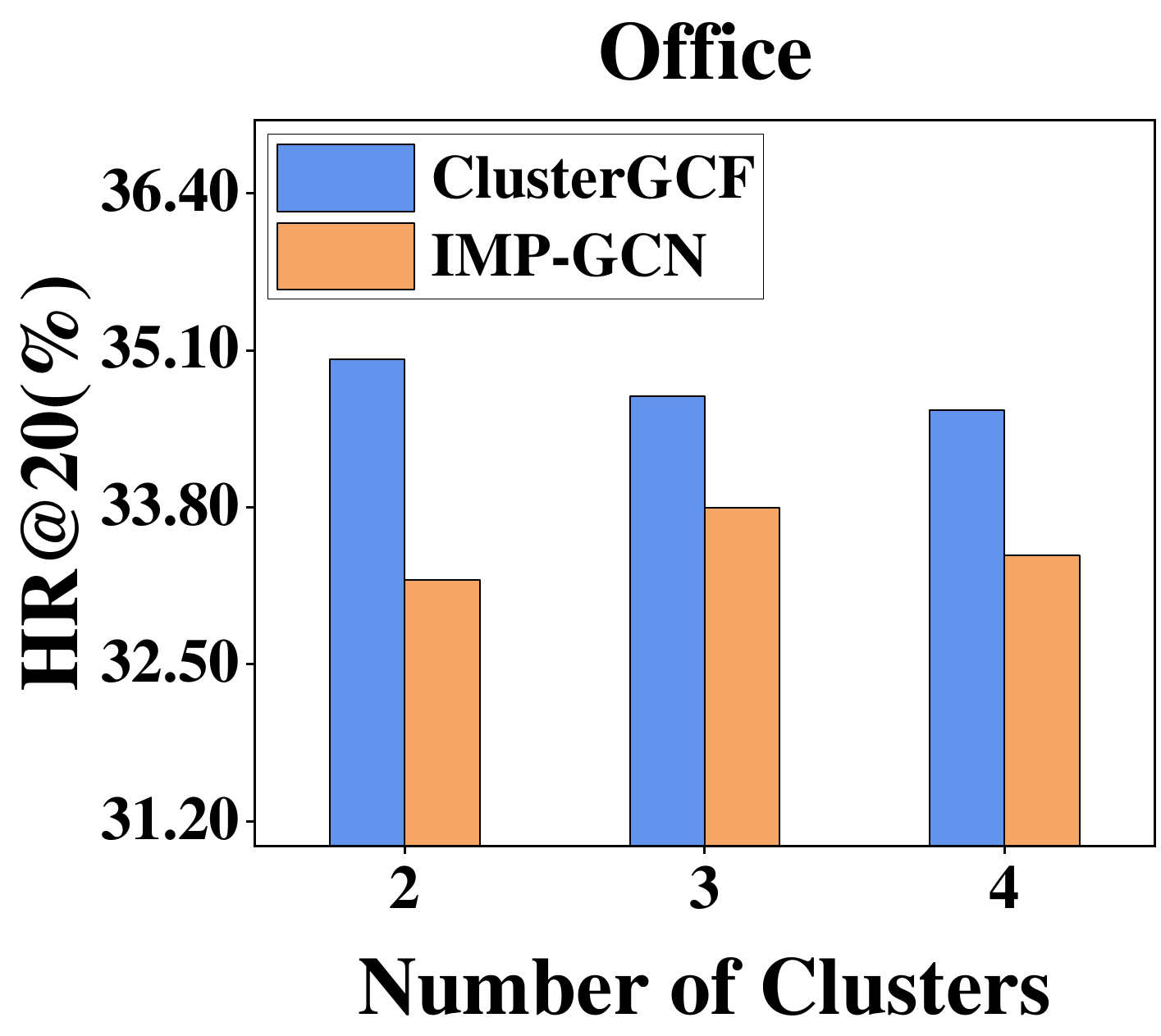}}
    {\includegraphics[width=0.24\linewidth]{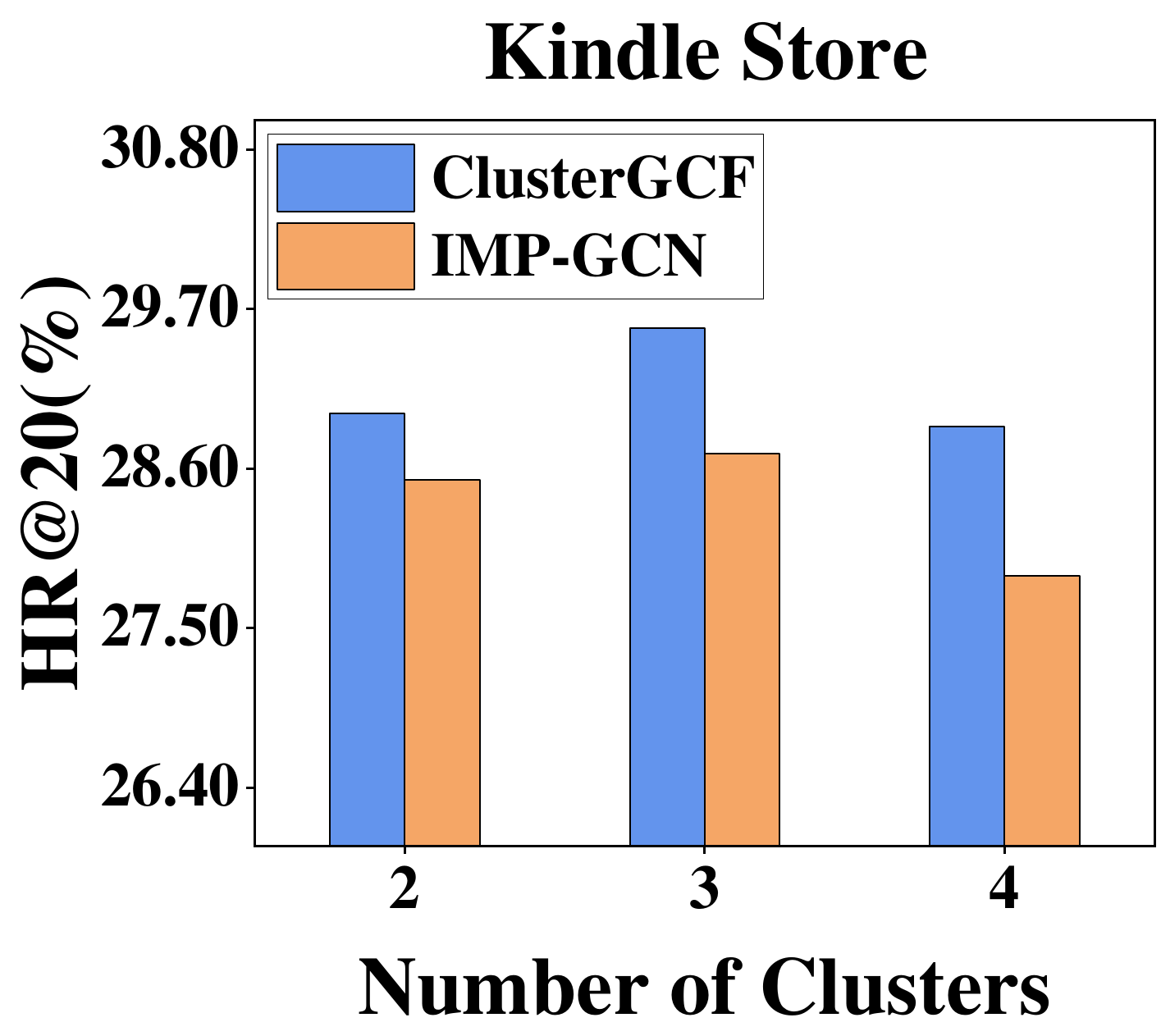}}
    {\includegraphics[width=0.24\linewidth]{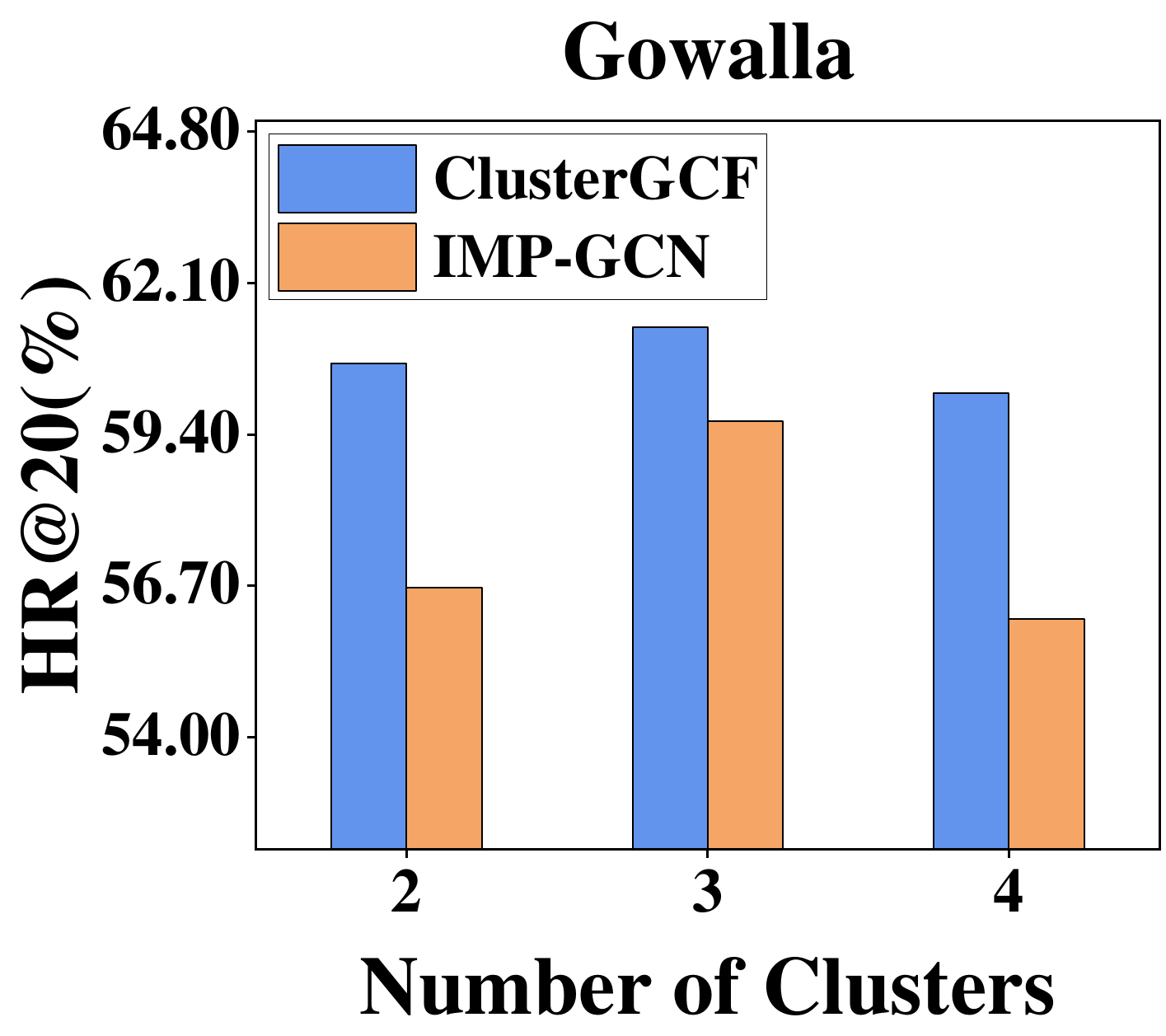}}
    {\includegraphics[width=0.24\linewidth]{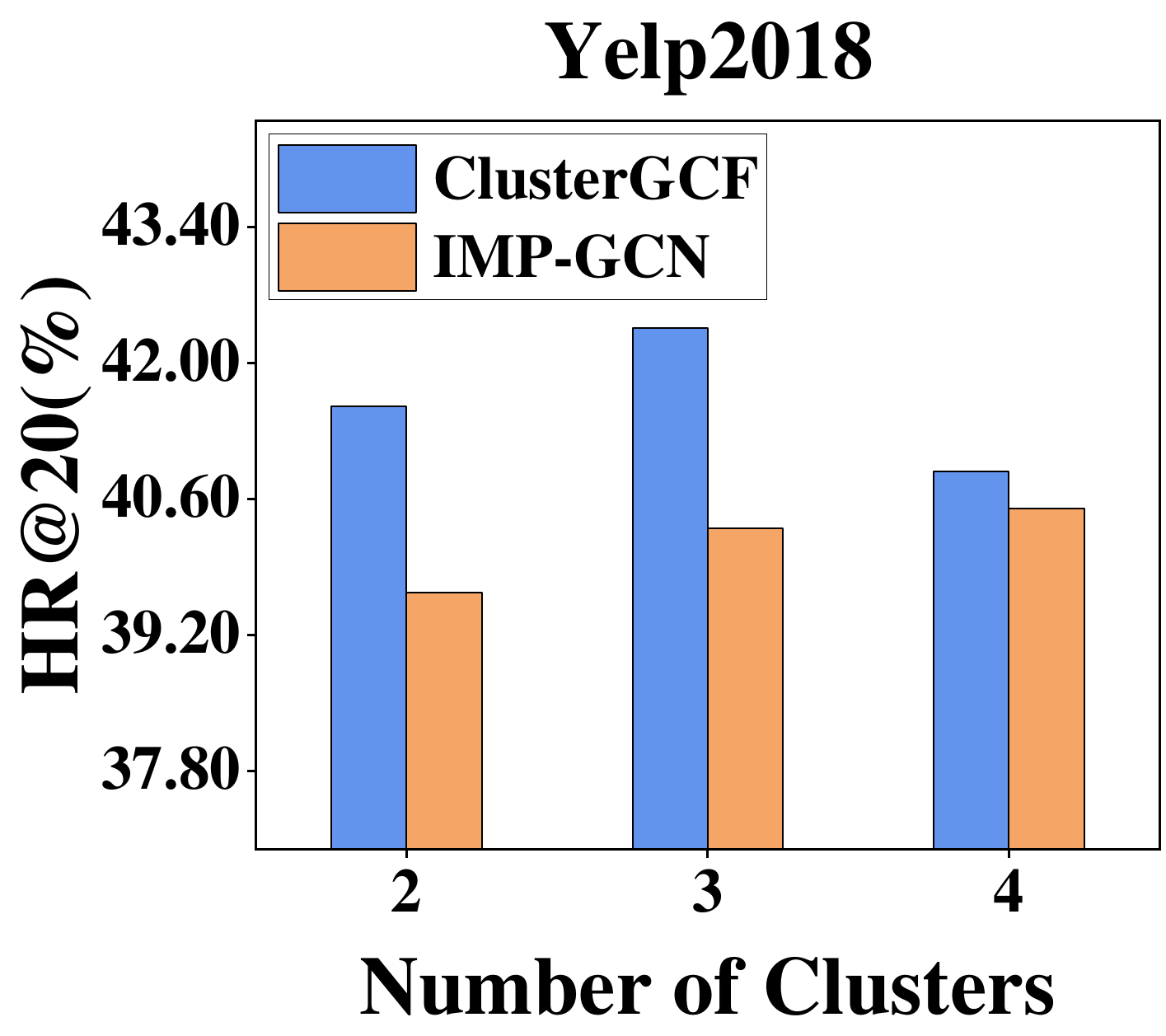}}\\
	{\includegraphics[width=0.24\linewidth]{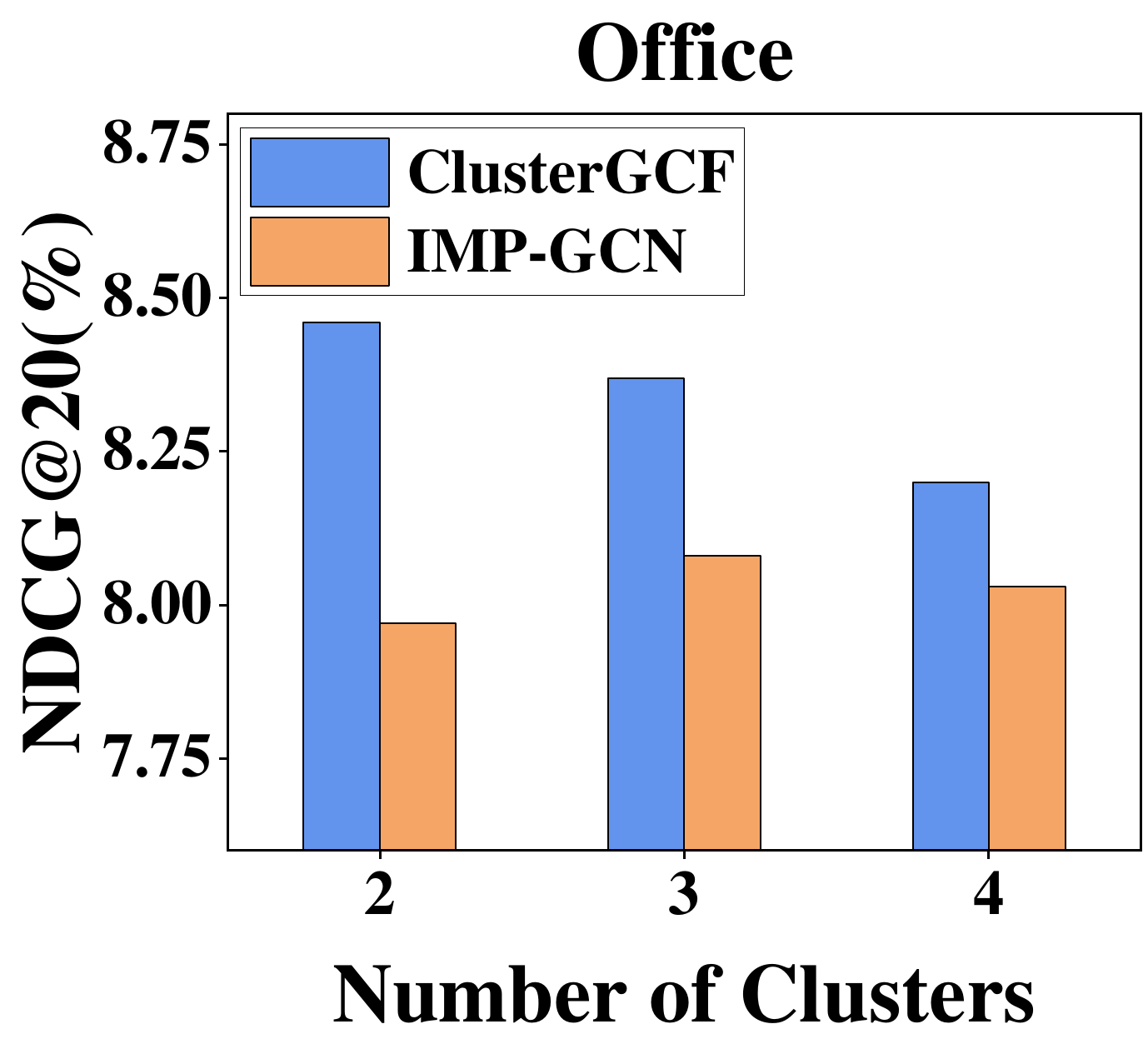}}
	{\includegraphics[width=0.24\linewidth]{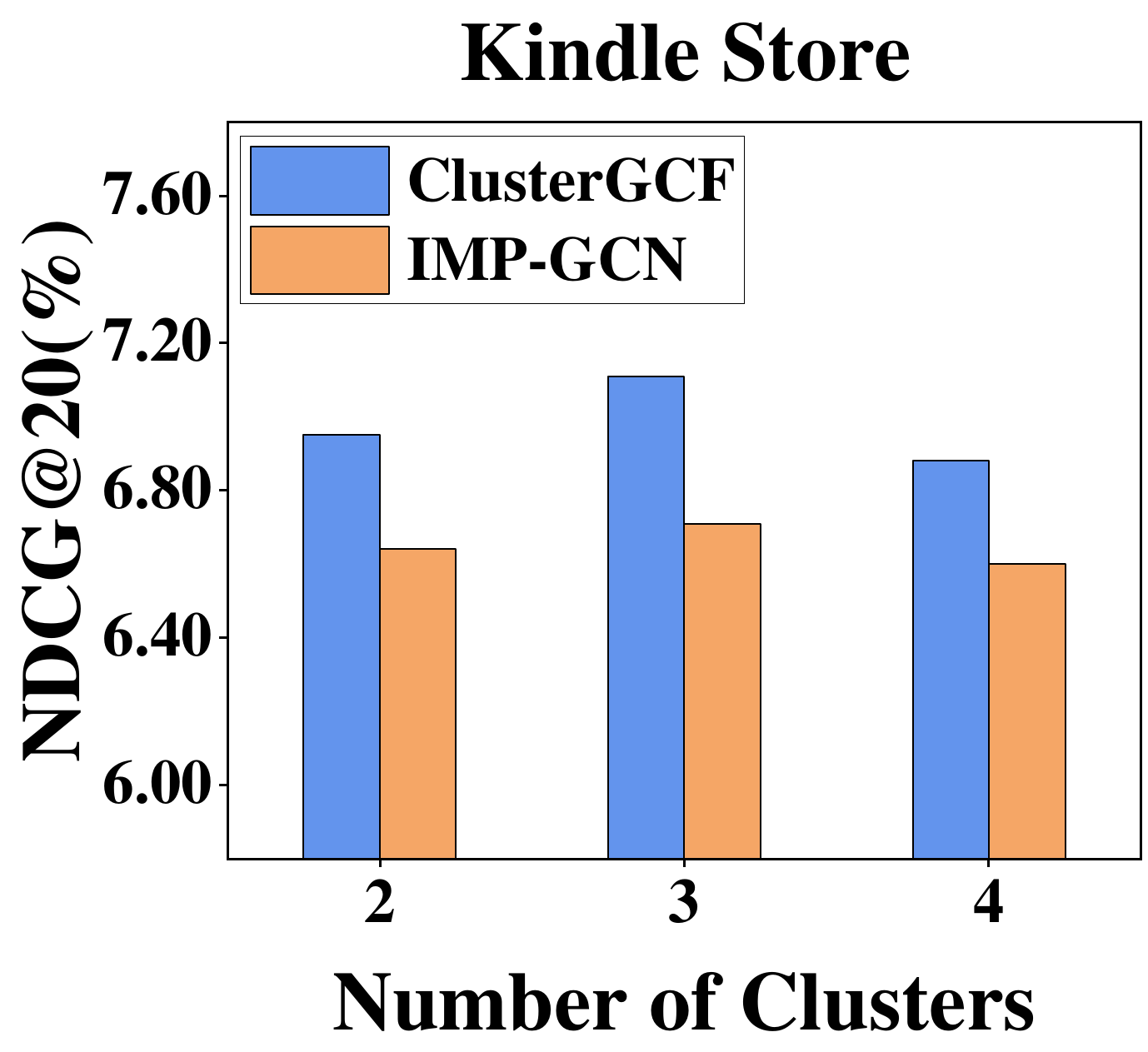}}
    {\includegraphics[width=0.24\linewidth]{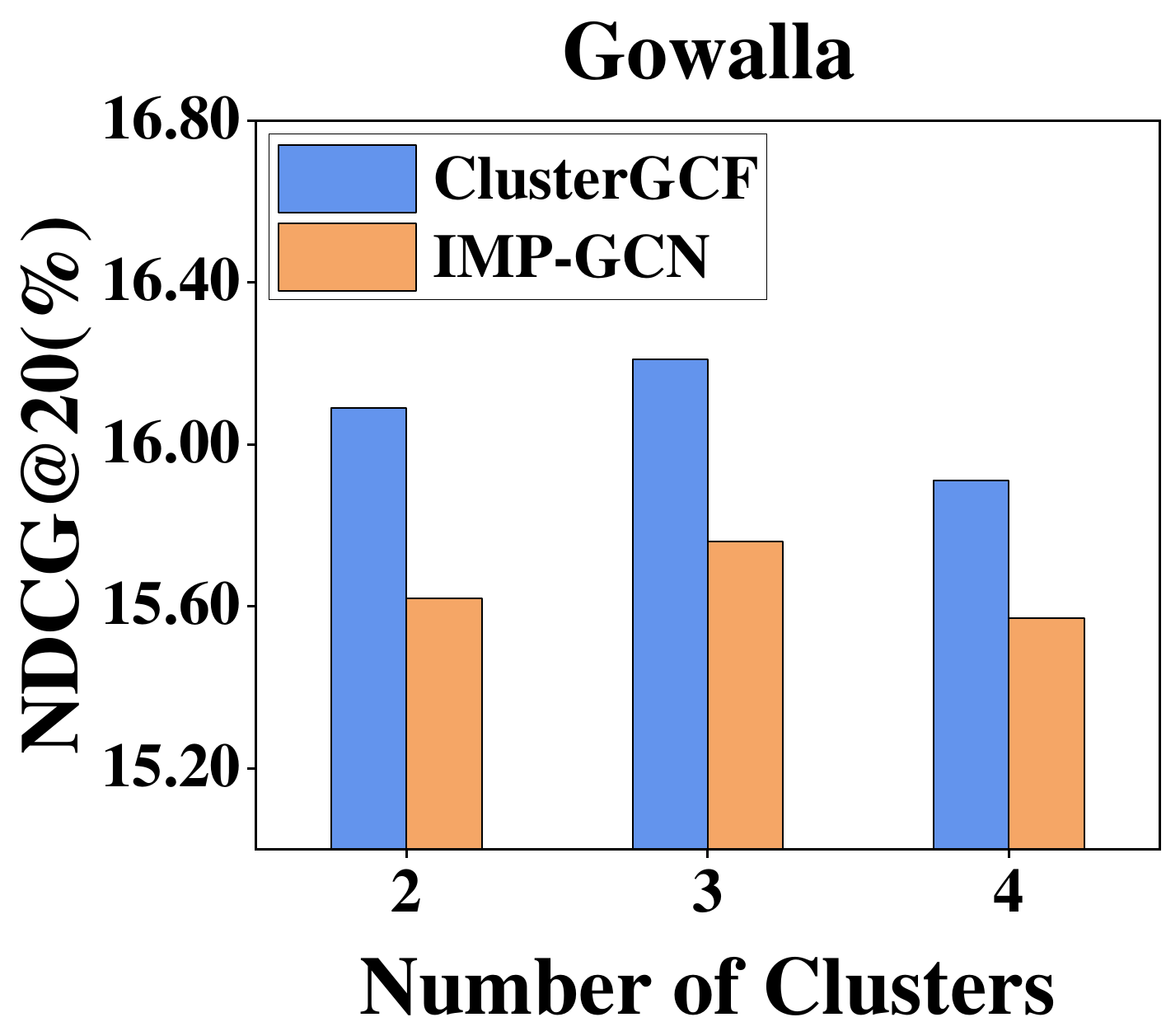}}
    {\includegraphics[width=0.24\linewidth]{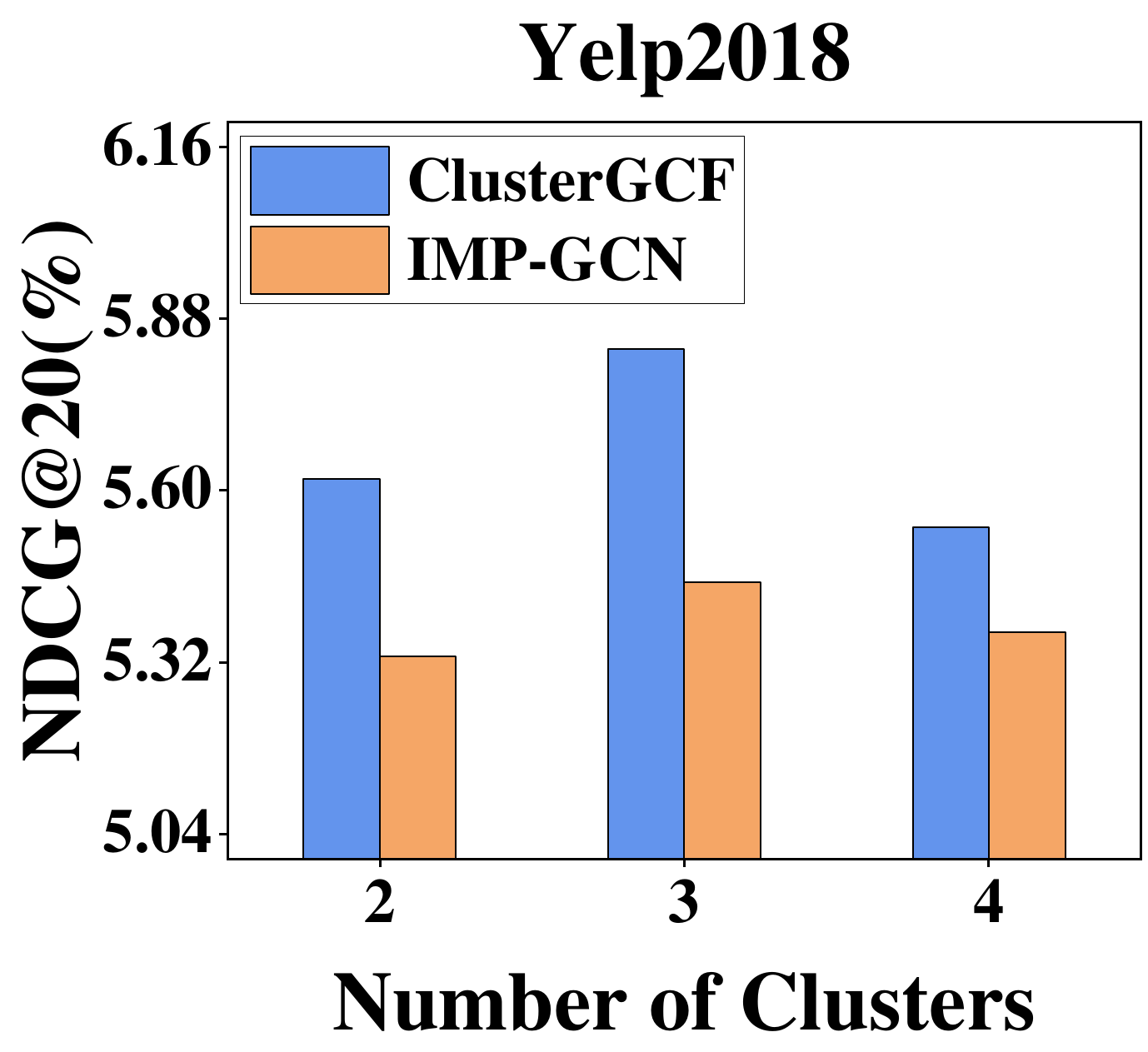}}
	\vspace{-0pt}
	\caption{Performance Comparison between ClusterGCF and IMP-GCN with different cluster numbers on all datasets.} 
	\label{fig:group_result}
\end{figure*}

In this section, we study the effects of cluster-specific graphs (numbers and the starting layer of graph convolution on the cluster-specific graphs) on the performance of our ClusterGCF. Besides, we also analyze the model Convergence of ClusterGCF to verify the effectiveness of employing cluster-specific graphs.
\subsubsection{Cluster-specific Graph Numbers}
To verify the effectiveness of our node clustering approach, we evaluated ClusterGCF and IMP-GCN across various numbers of clusters, specifically $\{2, 3, 4\}$, on all datasets. The experimental results are illustrated in Figure~\ref{fig:group_result}. 

From the results, we can see that ClusterGCF outperforms IMP-GCN across all cluster number settings. This can be attributed to several reasons: (1) \textbf{Item Node Clustering}. ClusterGCF offers an advantage by distinguishing item nodes among neighbors, which effectively reduces noise while improving valuable information from high-order neighbors. Specifically, IMP-GCN only uses users of common interest and their associated items to build subgraphs. In contrast, ClusterGCF builds cluster-specific graphs considering both the users with common interests and the items that appeal to them. (2) \textbf{Soft Clustering}. In ClusterGCF, the nodes (user nodes or item nodes) are assigned weights that denote the probabilities of their association with a particular cluster. This approach allows a node presence in multiple clusters. Thus, several cluster-specific graph might share some important nodes, enhancing the quality of information in the embedding propagation operation. (3) \textbf{Optimizable}. In comparison to IMP-GCN, the adopted node clustering method is optimized. The node clustering approach is co-trained with the recommendation model.

Another interesting finding is that ClusterGCF achieves its peak performance with 2 cluster-specific graphs on Office. This is because there is a trade-off in selecting the number of cluster-specific graphs~\cite{Liu2021IMP_GCN}. Having more cluster-specific graphs allows ClusterGCF to distinguish users and items at a more finer level, which enables it to distill information more effectively from high-order neighbors. However, it also reduces the valuable information from the neighbors in short-distance in embedding learning. Especially, IMP-GCN consistently underperforms ClusterGCF across all cluster-specific graph numbers, further highlighting the effectiveness of passing messages with considering users' multiple interests. 

\subsubsection{The Starting Layer of Cluster-based Graph Convolution}
To investigate the effect of the starting layer of cluster-based graph convolution, we devised the following two variants:
\begin{itemize} 
	\item \textbf{ClusterGCF$_{F}$}: This variant performs both the first- and high-order graph convolution on the cluster-specific graphs.
	\item \textbf{ClusterGCF$_{T}$}: In this variant, both the first- and second-order embedding propagation are conducted on the original user-item interaction graph.
\end{itemize}

Table~\ref{tab:F-T_results} presents the results of two variants as well as ClusterGCF.
The experimental results show that ClusterGCF$_{T}$ performs better performance than ClusterGCF$_{F}$. This indicates that the neighbors of short distance play a crucial role in the representation learning of the target node. In other words, the poor performance of ClusterGCF$_F$ is because distinguishing the first-order neighbors at a finer level may diminish the valuable information they provide. The result of ClusterGCF$_T$ also indicates the effectiveness of differentiating the high-order neighbor nodes. 
Across all datasets, ClusterGCF consistently outperforms both ClusterGCF$_F$ and ClusterGCF$_T$. This means that our ClusterGCF can achieve its peak performance on these three datasets when we perform second- and higher-order graph convolution on the cluster-specific graphs, as demonstrated by IMP-GCN~\cite{Liu2021IMP_GCN}. More specifically, the information from the second-order neighbors may introduce the noise information, resulting in sub-optimal performance.

\begin{table}[t]
	\caption{Performance comparison between ClusterGCF and its variants on four datasets. Please note that the values are presented as percentages, with the '\%' symbol omitted.} 
	\centering
	\resizebox{1.0\textwidth}{!}{
		\begin{tabular}{|l|ccc|ccc|ccc|ccc|} \hline
			Datasets	& \multicolumn{3}{c|}{Office} & \multicolumn{3}{c|}{Kindle Store} & \multicolumn{3}{c|}{Gowalla} & \multicolumn{3}{c|}{Yelp2018} 
            \\ \cline{2-4}  \cline{5-7} \cline{8-10} \cline{11-13}
			Metrics	& Recall	&HR &	NDCG	& Recall	&HR & NDCG	& Recall  &HR  &	NDCG	& Recall    &HR &	NDCG	\\ \hline \hline 
			ClusterGCF$_{F}$	&	13.78	& 34.65 &   8.25	& 10.94  &  28.91  & 6.8  &	18.74	& 60.44 &	15.92  	&	6.74 & 40.72	&   5.55 \\ 
			ClusterGCF$_{T}$ &	13.86 & 34.78	& 8.40 	& 	11.05 & 29.33	& 6.95  & 18.88  & 60.71 & 16.09	&	6.86 & 41.18	&   5.61  	\\  \hline \hline
			ClusterGCF	 &	\textbf{13.97}	& \textbf{35.02}	& \textbf{8.46}	&	\textbf{11.16}	&  \textbf{29.57}	& \textbf{7.11} & \textbf{19.07}	& \textbf{61.31}	 & \textbf{16.21}	& \textbf{7.06}	& \textbf{42.36}	 & \textbf{5.83}\\ \hline
	\end{tabular}}
	\label{tab:F-T_results}
	\vspace{0pt}
\end{table}

\begin{figure*}[t]
    \centering
	\hspace{0.0cm}
    {\includegraphics[width=0.3\linewidth]{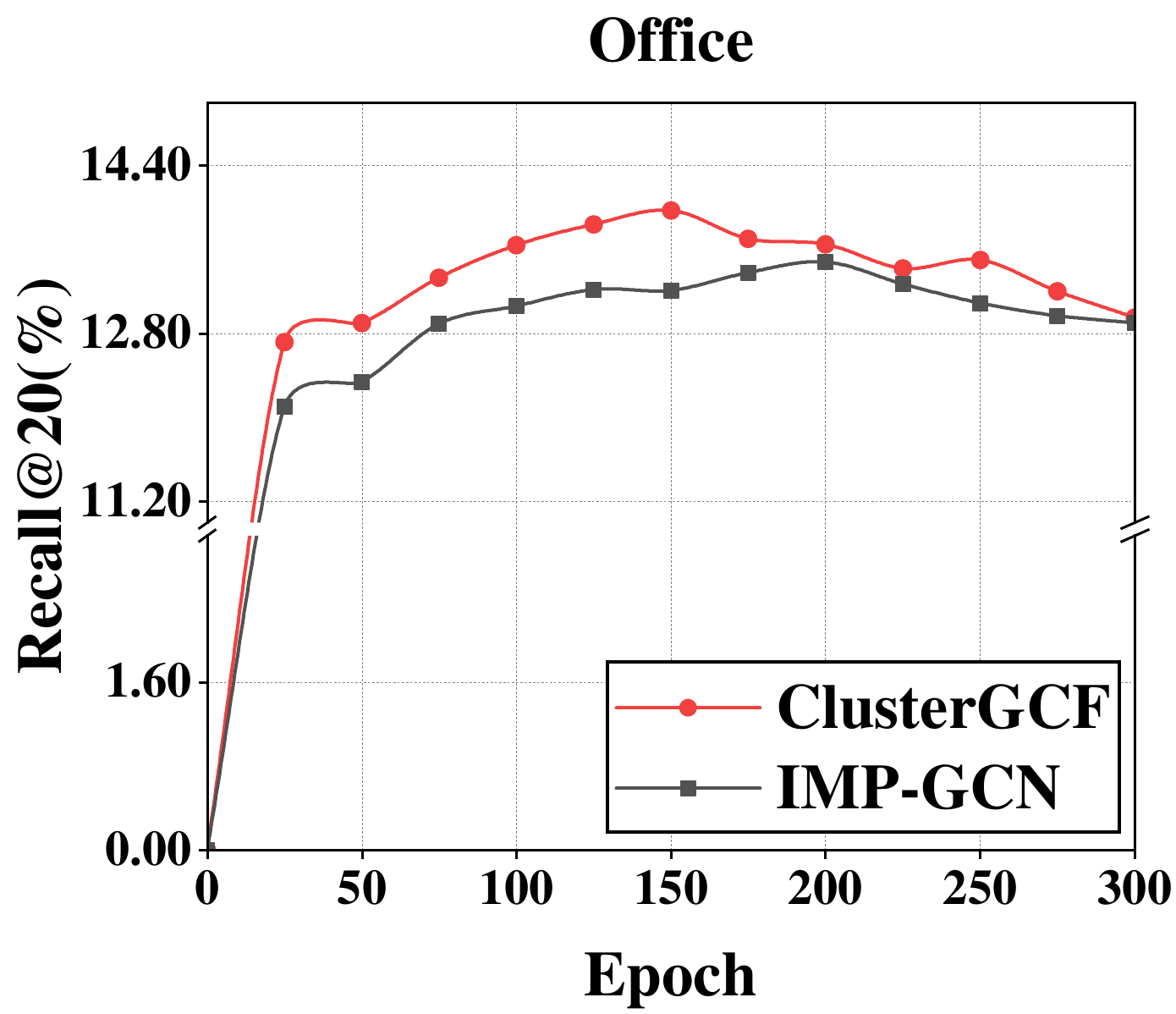}}\hspace{40pt}
    {\includegraphics[width=0.3\linewidth]{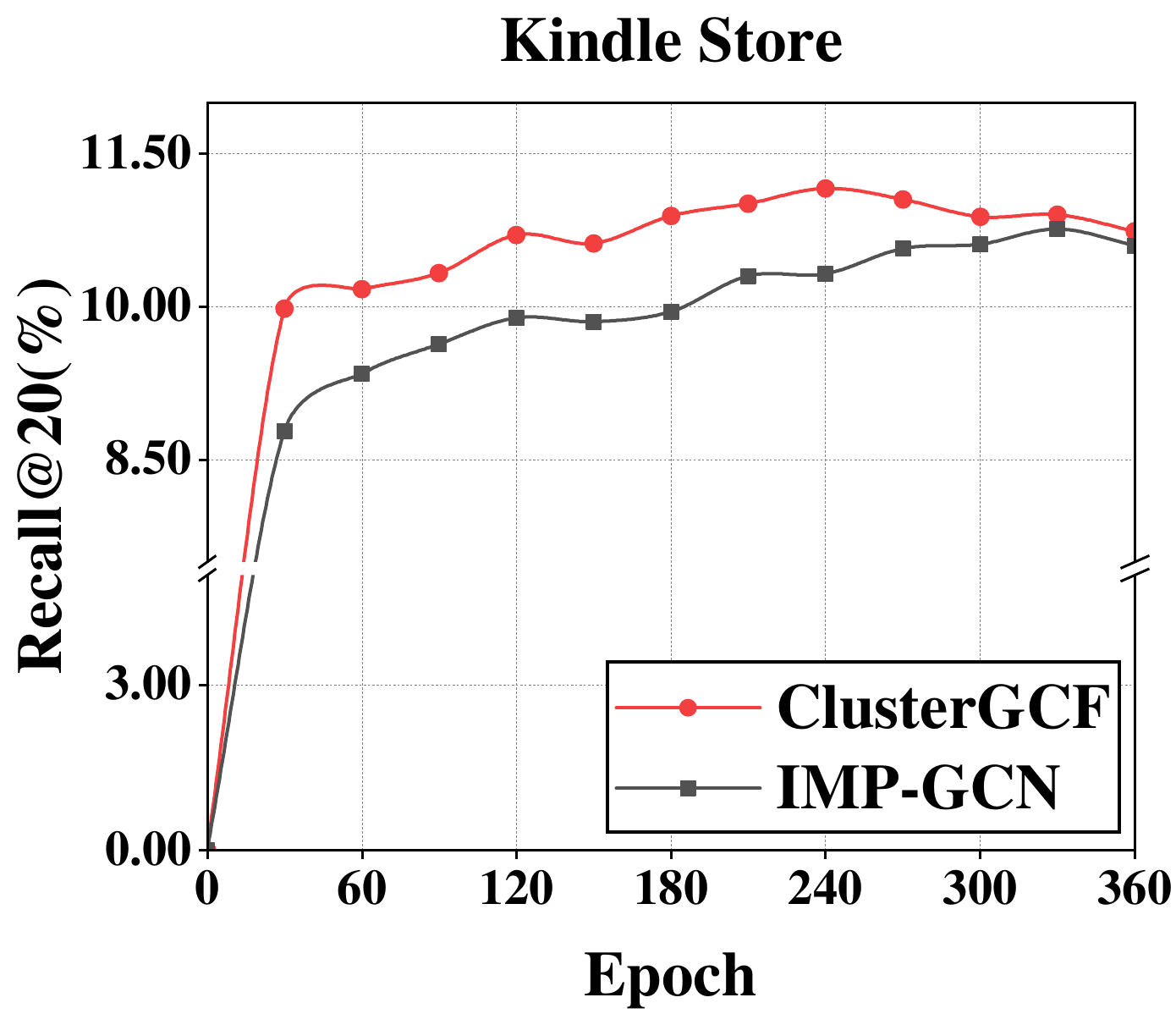}}\\
    
    {\includegraphics[width=0.3\linewidth]{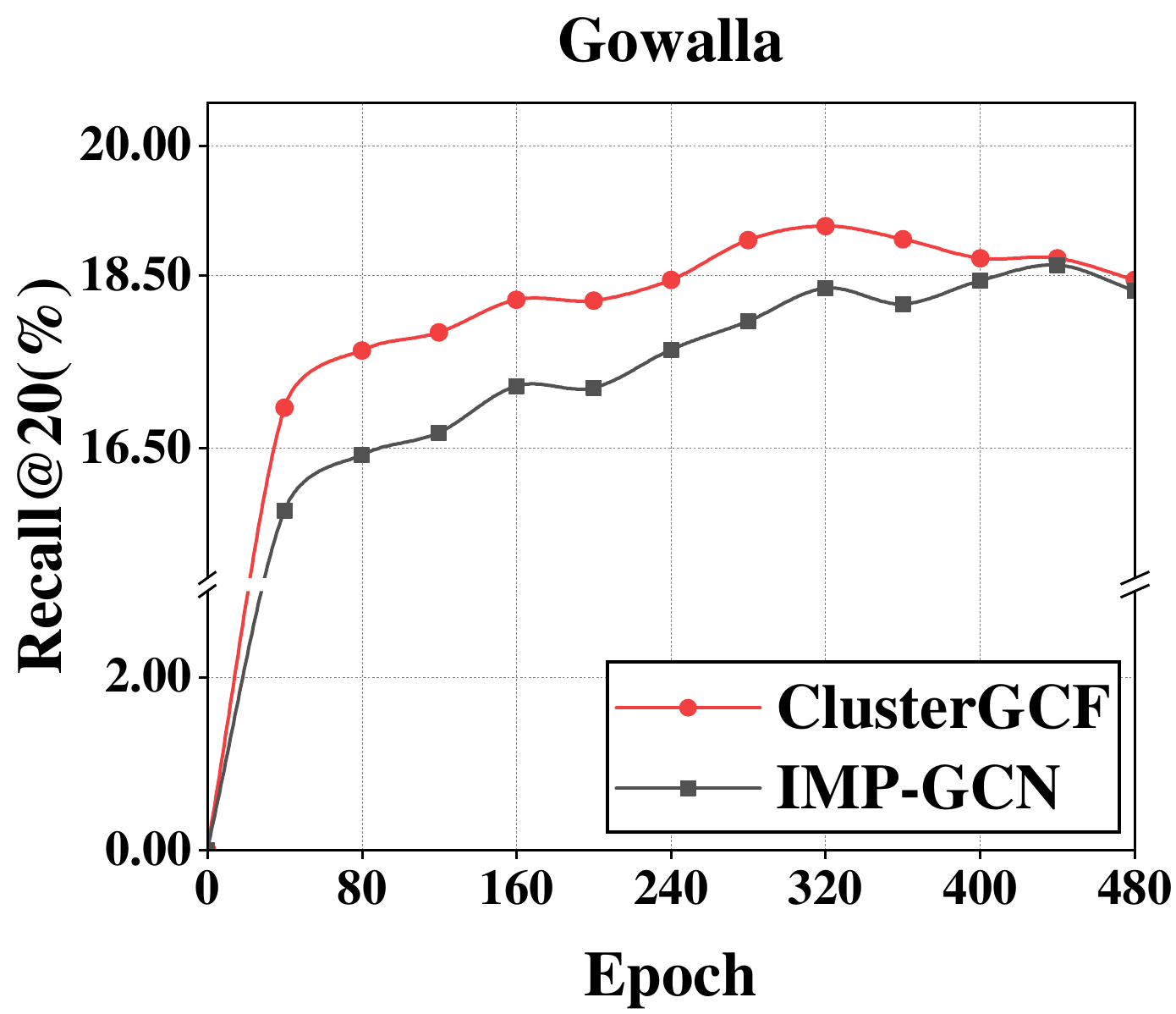}}\hspace{40pt}
    {\includegraphics[width=0.3\linewidth]{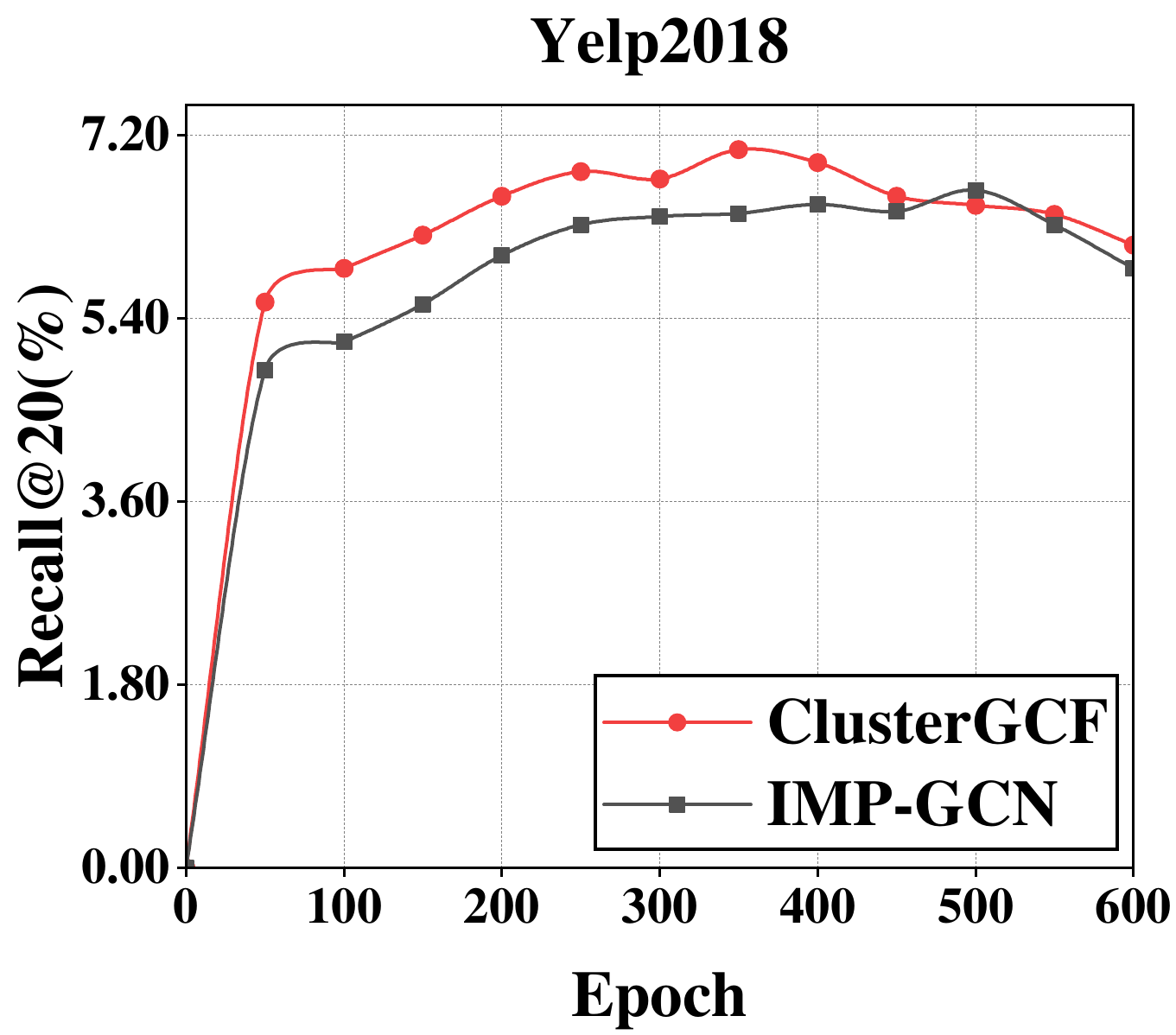}}\\
	\vspace{0pt}
	\caption{Performance evaluation for each epoch of ClusterGCF and IMP-GCN.}
	\vspace{0pt}
	\label{fig:test-epoch}
\end{figure*}

\subsubsection{Model Convergence.}

We analyze the convergence of ClusterGCF and IMP-GCN based on their performance in terms of Recall over four datasets. 

As illustrated in Figure~\ref{fig:test-epoch}, ClusterGCF exhibits faster convergences than IMP-GCN. For example, ClusterGCF reaches peak performance at 320 epochs, while IMP-GCN requires 360 epochs on the Kindle Store dataset. This can be attributed to the soft clustering method employed in ClusterGCF. Soft clustering provides more flexibility for the model. When the model tries to find the optimal classification or clusters, it does not need to forcefully assign a node to a specific cluster but can instead assign a probability to it. In other words, the model can adjust its decision boundaries more smoothly, accelerating the learning process. Due to the smooth nature of soft clustering, the model might encounter fewer local optima and traps. Hence, the optimization algorithm could more easily find a good solution. In summary, soft clustering offers smoother and more informative feedback, aiding the model in learning and converging more rapidly.

\begin{figure}[t]
	\centering
	{\includegraphics[width=0.3\linewidth]{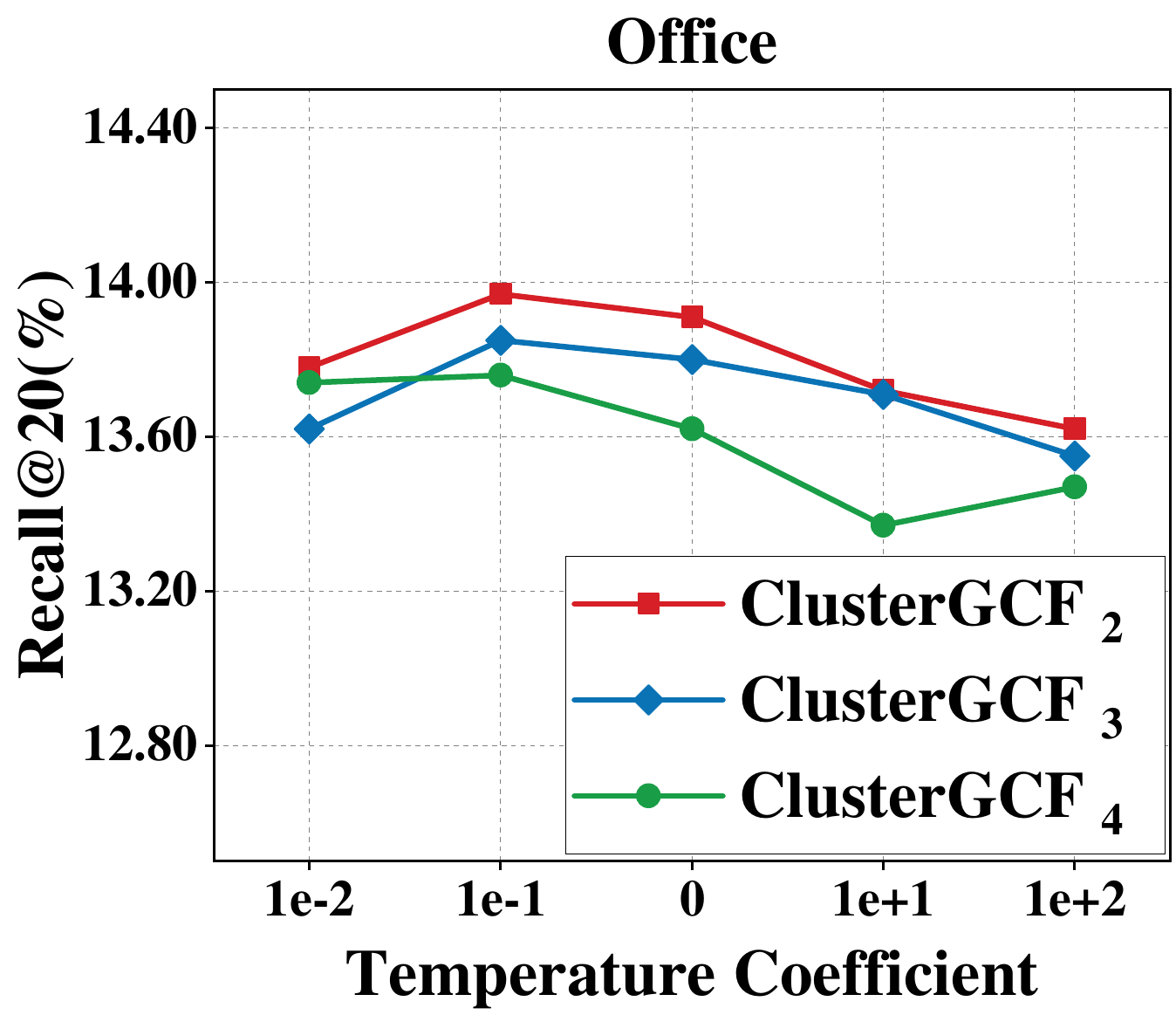}}\hspace{40pt}
    {\includegraphics[width=0.3\linewidth]{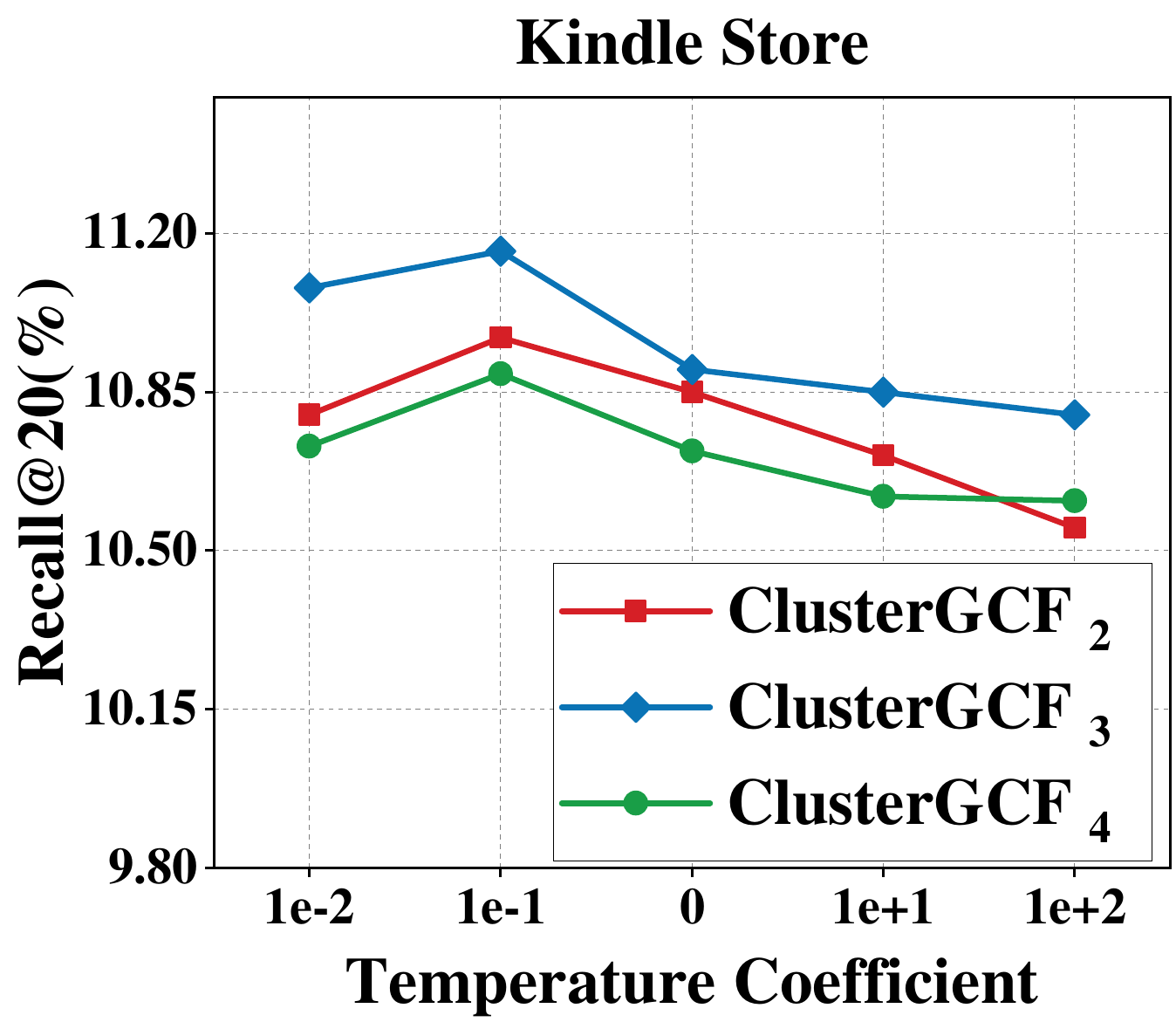}}\\
    {\includegraphics[width=0.3\linewidth]{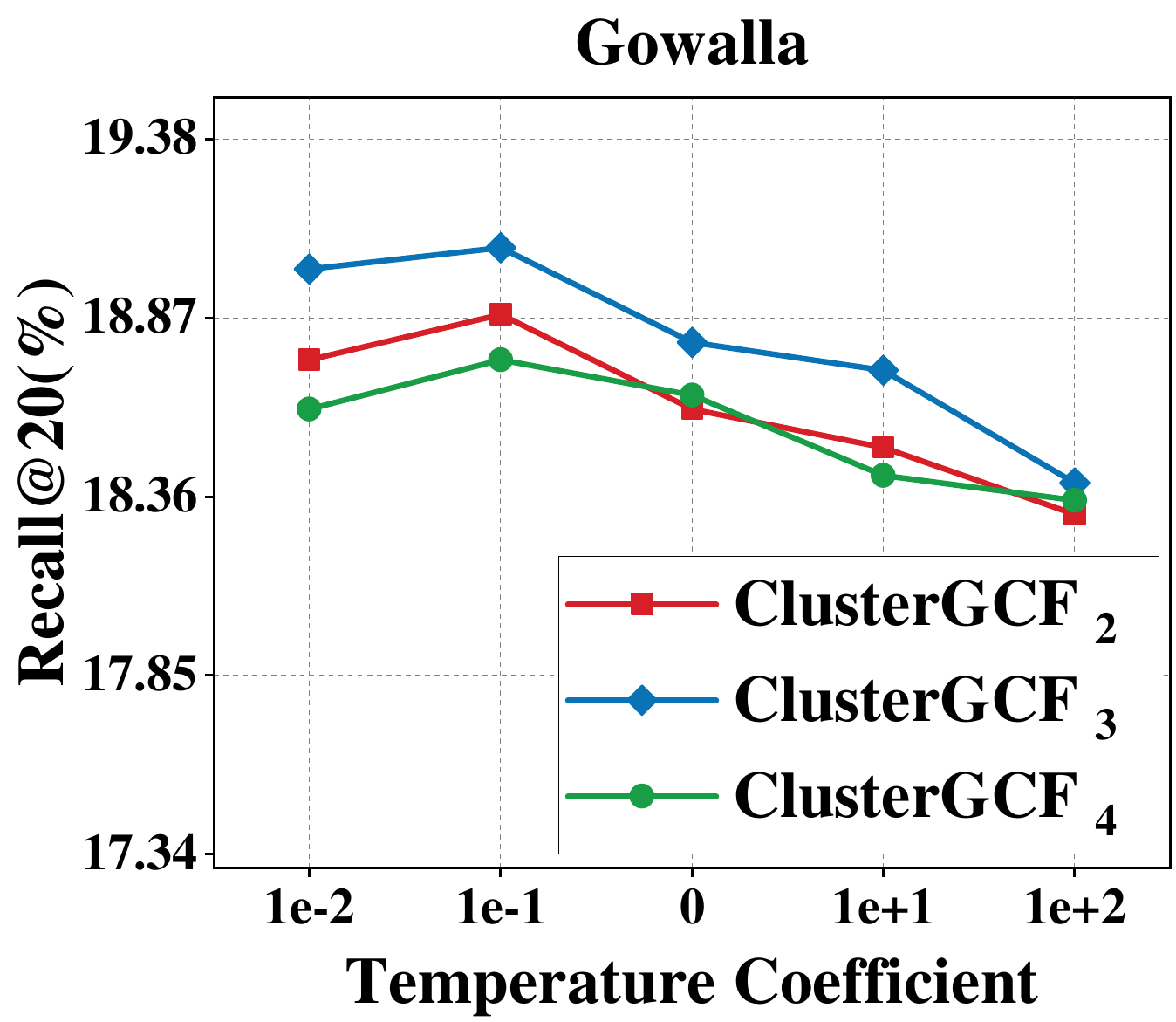}}\hspace{40pt}
	{\includegraphics[width=0.3\linewidth]{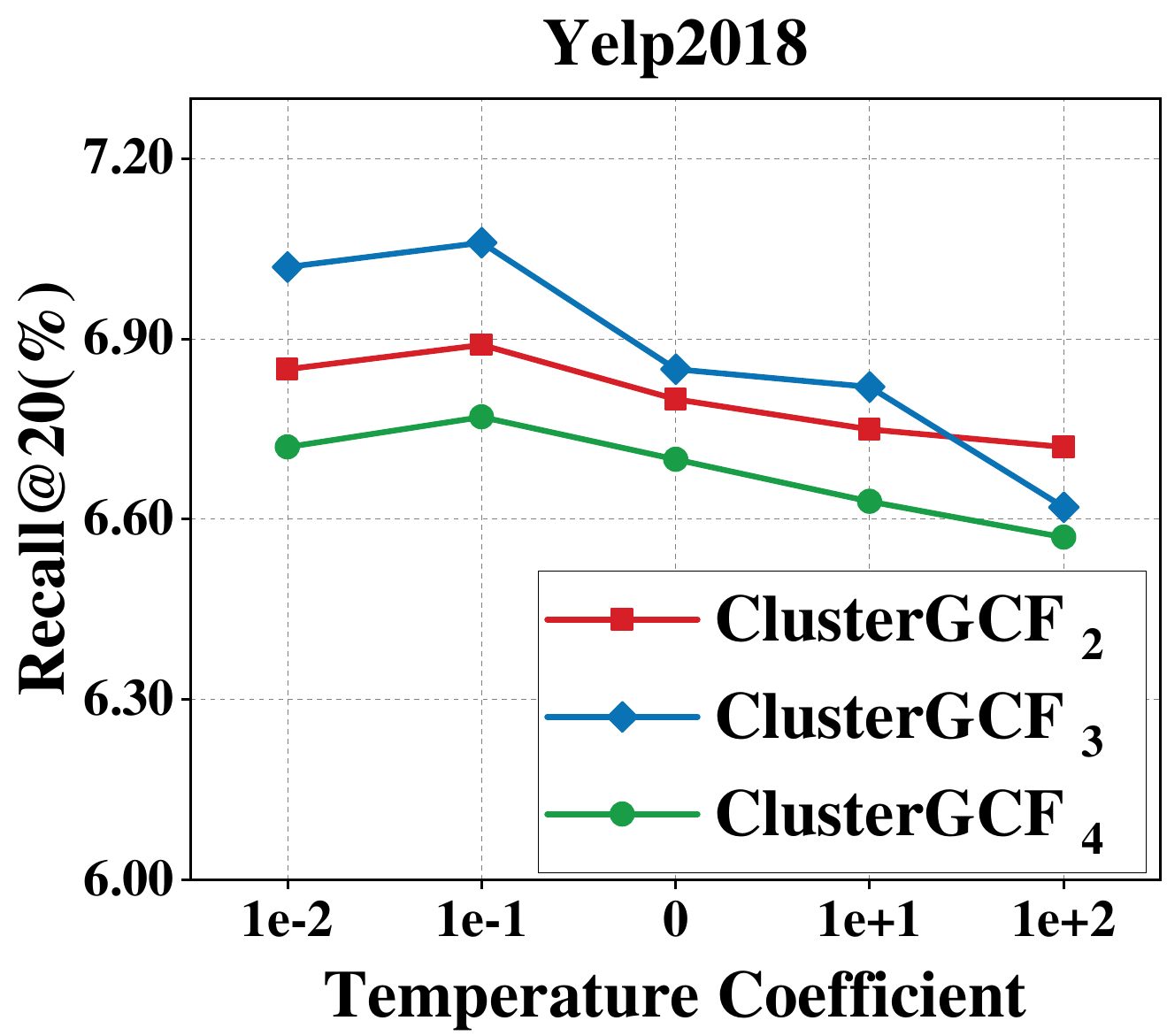}}\\
	\vspace{-0pt}
	\caption{Effect of Temperature Coefficient on the Performance of ClusterGCF Models. ClusterGCF$_2$, ClusterGCF$_3$ and ClusterGCF$_4$ represent ClusterGCF with 2, 3, and 4 cluster-specific graphs,respectively.}
	\vspace{0pt}
	\label{fig: temperature Coefficient}
\end{figure}

\subsection{Effect of Node Clustering (RQ4)}
In this section, we examine the impact of the temperature coefficient on node clustering. Additionally, we investigate the influence of node type in clustering and the effect of the clustering method on cluster-specific graph construction.

\subsubsection{Temperature Coefficient.}
In our ClusterGCF model, the Gumbel-Softmax technique plays a pivotal role in facilitating soft clustering for user and item nodes. A key component in this method is the temperature coefficient, which controls the smoothness of the output. Lower values make the output closer to one-hot encoded vectors (hard classification), while higher values make the output more uniform (soft classification). To investigate the impact of the temperature coefficient on ClusterGCF, we performed experiments by varying the value of the temperature coefficient in the range of $\{1e^{-2},1e^{-1},\cdots,1e^{2}\}$. Then, we evaluated the model's performance with different cluster-specific graph numbers on different datasets in terms of \emph{Recall}~\footnote{The performance of ClusterGCF with respect to \emph{HR} and \emph{NDCG} show a similar trend as \emph{Recall}.}. 

Figure ~\ref{fig: temperature Coefficient} illustrates the performance of ClusterGCF on all datasets. From the results, with increasing of the temperature coefficient, there was a noticeable improvement in performance. This is because more valuable information can be passed from high-order neighbors. The clusterGCF reached its peak performance when the temperature coefficients were set to $1e^{-1}$. After that, further amplification in the coefficient beyond this threshold led to a performance drop-off. This is because higher values make the clustering results uniform. Thus, ClusterGCF cannot differentiate nodes among high-order neighbors, resulting in suboptimal performance. 

\subsubsection{Node Types.}
To investigate the effect of node type involved in node clustering for ClusterGCF, we devised the following two variants:
\begin{itemize}
	\item \textbf{ClusterGCF$_{user}$}: This variant solely classifies user nodes into various clusters based on their shared interest. 
	\item \textbf{ClusterGCF$_{item}$}: This variant solely categorized item nodes into distinct clusters based on their common characteristics.
\end{itemize}

\begin{table}[t]
	\caption{Performance of IMP-GCN, ClusterGCF and its variants over four datasets. Notice that the values are reported by percentage with '\%' omitted.} 
	\centering
	\resizebox{1.0\textwidth}{!}{
		\begin{tabular}{|l|ccc|ccc|ccc|ccc|} \hline
			Datasets	& \multicolumn{3}{c|}{Office} & \multicolumn{3}{c|}{Kindle Store} & \multicolumn{3}{c|}{Gowalla} & \multicolumn{3}{c|}{Yelp2018} 
            \\ \cline{2-4}  \cline{5-7} \cline{8-10} \cline{11-13}
			Metrics	& Recall	&HR &	NDCG	& Recall	&HR & NDCG	& Recall  &HR  &	NDCG	& Recall    &HR &	NDCG	\\ \hline \hline 
            IMP-GCN	&	13.48 &	33.84 &   8.11	& 10.77   & 28.61 & 6.75  &	18.62	& 59.77 &	15.76  	&	6.66	& 40.43 &	5.45  \\
            ClusterGCF$_{user}$ &	13.88 & 34.95	& 8.38 	& 	11.03 & 29.12	& 7.03  & 18.97  & 61.07 & 16.09	&	6.94 & 41.85	&   5.67  	\\
   		ClusterGCF$_{item}$ &	13.72 & 34.51	& 8.22 	& 	10.84 & 28.77	& 6.82  & 18.85  & 60.60 & 16.01	&	6.82 & 40.04	&   5.50  	\\  
              \hline \hline
			ClusterGCF	 &	\textbf{13.97}	& \textbf{35.02}	& \textbf{8.46}	&	\textbf{11.16}	&  \textbf{29.57}	& \textbf{7.11} & \textbf{19.07}	& \textbf{61.31}	 & \textbf{16.21}	& \textbf{7.06}	& \textbf{42.36}	 & \textbf{5.83}\\ \hline
	\end{tabular}}
	\label{tab:nodetype}
	\vspace{0pt}
\end{table}
In this section, we investigate different node types to validate the effectiveness of node clustering within ClusterGCF model. Specifically, the ClusterGCF$_{user}$ focuses on clustering user nodes, while ClusterGCF$_{item}$ is dedicated to item nodes. We present a comparative analysis of our ClusterGCF, its two specific variants, and IMP-GCN. The experimental results are detailed in Table ~\ref{tab:nodetype}. These results offer several intriguing insights.

First, ClusterGCF$_{user}$ achieved better performance than IMP-GCN. This is because ClusterGCF$_{user}$ benefits from the adopted clustering approach which is optimizable and it clusters user nodes in a soft manner. As a result, the target node can obtain more valuable information from high-order neighbors in representation learning. Second, ClusterGCF$_{item}$ also performs better than IMP-GCN, indicating the effectiveness of item node clustering within ClusterGCF. Lastly, ClusterGCF consistently outperforms the ClusterGCF$_{user}$ and ClusterGCF$_{item}$ across all datasets. This demonstrates the robustness and effectiveness of the node clustering approach in ClusterGCF.

\subsection{Visualization (RQ5)}
\begin{figure}[t]
	\centering
    \subfloat[2 layers]{\includegraphics[width=1.8in]{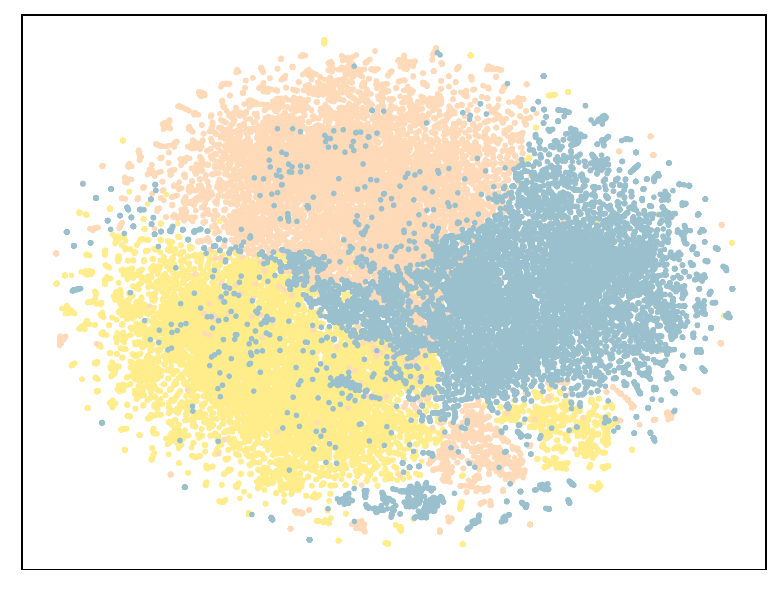}}
    \subfloat[3 layers]{\includegraphics[width=1.8in]{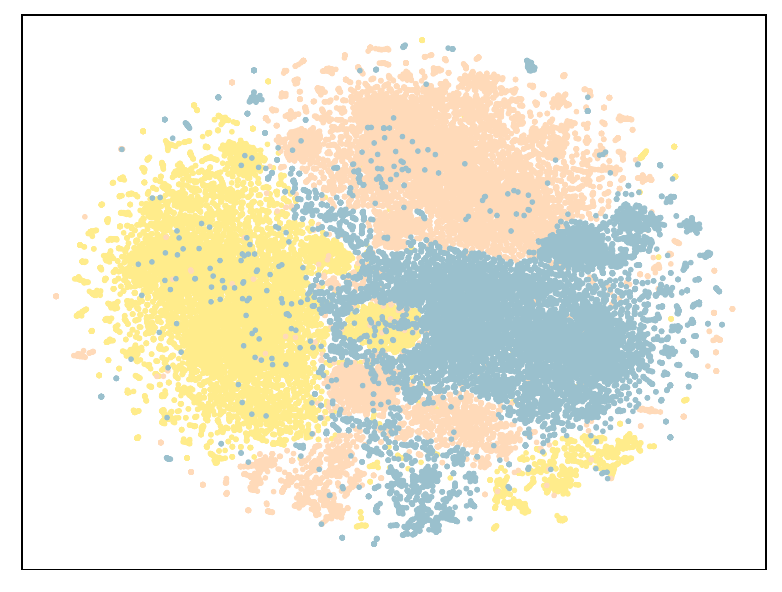}}
    \subfloat[4 layers]{\includegraphics[width=1.8in]{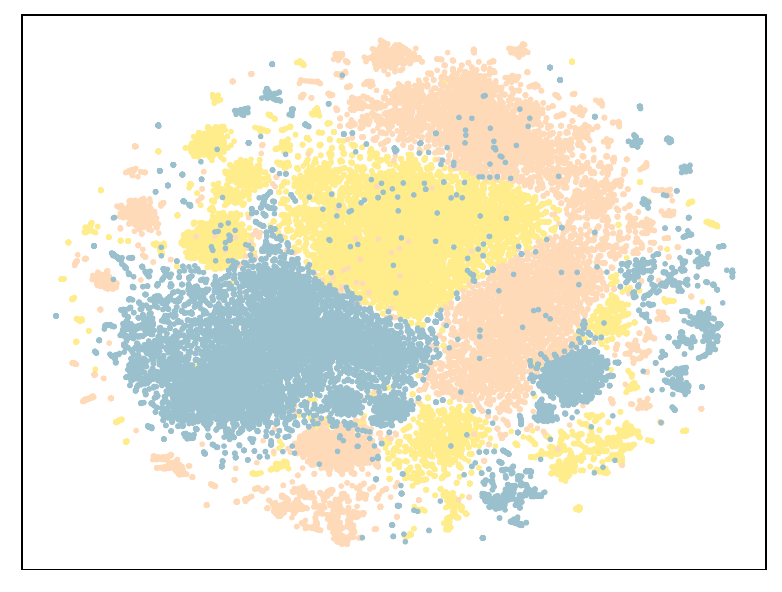}}\\
    \subfloat[5 layers]{\includegraphics[width=1.8in]{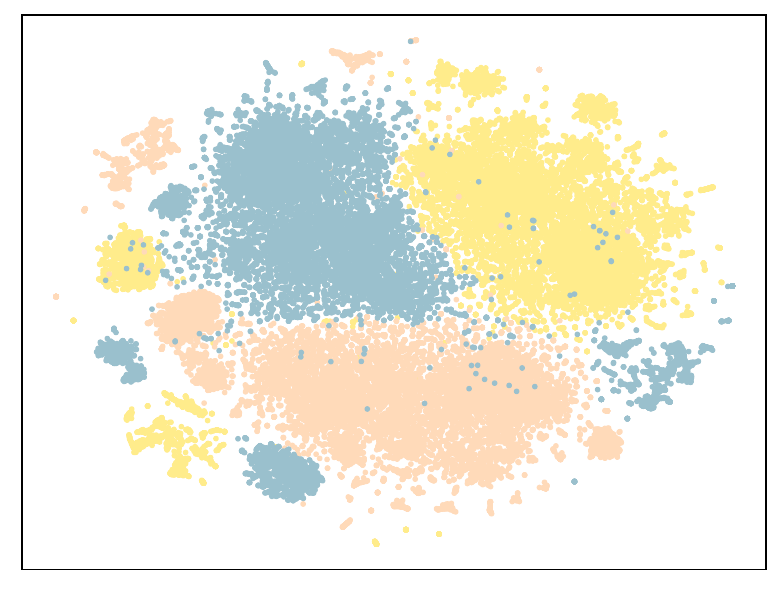}}
    \subfloat[6 layers]{\includegraphics[width=1.8in]{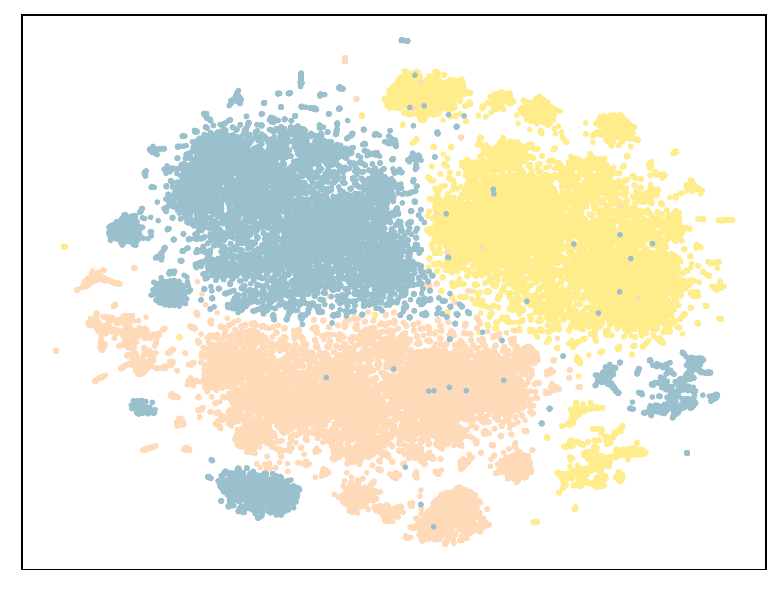}}
    \subfloat[7 layers]{\includegraphics[width=1.8in]{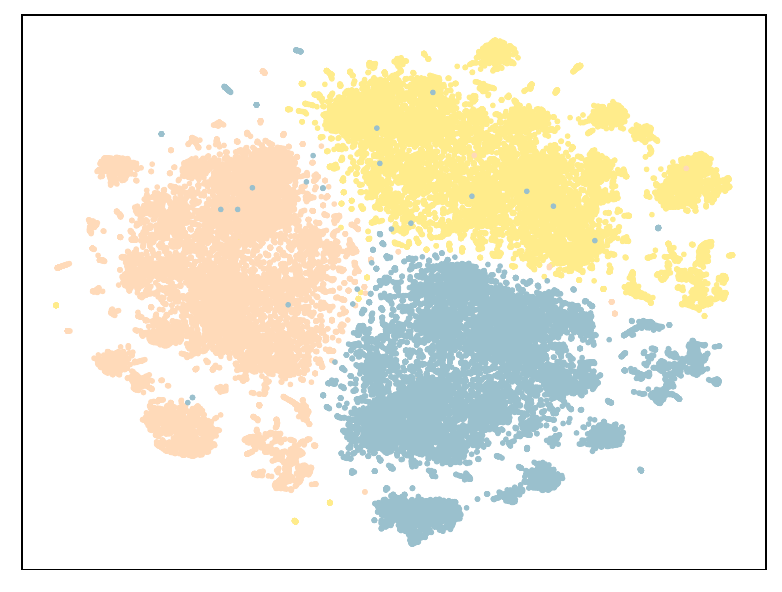}}
	\vspace{0pt}
	\caption{Visualization of propagated embeddings within different graph convolution layers for the target user nodes, where each color corresponds to a cluster.}
	\vspace{0pt}
	\label{fig: Visualization experiment}
\end{figure}
To further analyze the effect of ClusterGCF on representation learning within different graph convolution layers, we visualized the embeddings propagated from neighbor nodes of varying orders. Specifically, we first classify the user nodes into three clusters using the soft clustering method over the Kindle Store. Then, we visualized the embedding propagated from high-order neighbor nodes, using the t-SNE~\cite{vanDerMaaten2008} in Figure ~\ref{fig: Visualization experiment}. The figures show the embedding of neighbors of different orders, ranging from 2 to 7. Each dot in the figures denoted the embedding propagated from a user node, while dots of the same color across all figures denote the same cluster of users. Figure ~\ref{fig: Visualization experiment} shows the following interesting observations.

As graph convolution layers increase, the embedding vectors derived from high-order neighbors exhibit growing independence across different clusters. For instance, at the 2-order level, while embeddings within each cluster manifest specific aggregation tendencies, they are still comparatively scattered. With the addition of more layers, there's an evident shift in the node embeddings within each cluster. By the 7th layer, the dispersion notably diminishes, resulting in embeddings that display heightened independence. The representation learning in ClusterGCF is enhanced by these independent embeddings from high-order neighbors, as such independent representations have proven to be more robust against complex variations~\cite{ma2019learning}.

\subsection{Case Study}

\begin{figure}[t]
	\centering
\includegraphics[width=0.5\linewidth]{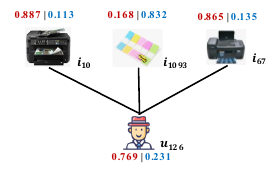}
	\caption{An example of high-order neighbor nodes in two cluster-specific graphs. The weights associated with each node are marked in red (cluster 1) and blue (cluster 2).}
	\vspace{0pt}
	\label{fig:casestudy}
\end{figure}

As shown in Fig.~\ref{fig:casestudy}, we conducted a case study to demonstrate the effectiveness of our proposed method in filtering out noisy neighborhoods. We choose a user node $u_{126}$ at the second graph layer and the interacted three item nodes $i_{10}$, $i_{67}$ and $i_{1093}$ on Office. The weights associated with each node within two cluster-specific graphs are marked in red (cluster 1) and blue (cluster 2). Although the user $u_{126}$ interacted with three items, two of them ($i_{10}$ and $i_{67}$) are assigned greater weights in cluster 1, while the other item ($i_{1093}$) is assigned a greater weight in cluster 2. Consequently, during the embedding propagation process, the influence of this node’s information is reduced in both cluster-specific graphs, demonstrating that our method can effectively filter out noisy information.

\section{Conclusion and Future work}

In this work, we argued that existing GCN-based methods are inadequate for obtaining reliable and adequate information from high-order neighboring nodes since the multiple interests of users are not properly considered. To alleviate this issue, we proposed the ClusterGCF model that performs high-order graph convolution inside cluster-specific graphs constructed by exploiting common interests among users and their multiple interests. To group both user and item nodes into multiple clusters, we introduced a soft node clustering method, which is unsupervised and optimizable, to estimate the probability assigned to the node for each cluster.  
The cluster-specific graphs are constructed relying on the topology of the user-item interaction graph and the clustering results.
Since the embeddings of nodes are propagated within cluster-specific graphs, ClusterGCF can filter out the negative information while capturing more valuable information from high-order neighboring nodes. 
Besides, the proposed clusterGCF maintains the uniqueness of node embeddings while stacking more graph convolution layers because it can benefit from higher-order neighboring nodes in cluster-specific graphs. Experiments on four datasets demonstrate that ClusterGCF can achieve state-of-the-art performance. The superior performance of ClusterGCF indicates the importance of distinguishing high-order neighbors in enhancing representation learning and demonstrates the effectiveness of the proposed soft node clustering approach.

This work attempts to exploit the common interest between users and the multiple interests of a user by grouping the users and items in the high-order graph convolution operation, limited to the unsupervised clustering approach and recommendation scenario. Specifically, there are many other clustering approaches, including using supervised information or additional knowledge. 
Furthermore, the cluster-based graph convolution operation can be widely applied in various tasks, such as social network analysis and information retrieval. We expect the potential of the cluster-based GCNs to be further explored towards other tasks.


\bibliographystyle{ACM-Reference-Format}
\bibliography{ClusterGCF}

\end{document}